\documentclass{article}
\usepackage{amssymb, amsmath}
\numberwithin{equation}{section}
\newtheorem{definition}{Definition}[section]
\newtheorem{theorem}[definition]{Theorem}
\newtheorem{proposition}[definition]{Proposition}
\newtheorem{lemma}[definition]{Lemma}

\def\supp{{\rm supp}}

\def\max{{\rm max}}

\begin{document}

\title{On surface-symmetric spacetimes with collisionless and charged 
matter}

\author{Sophonie Blaise Tchapnda}
\date{}
\maketitle{}
\begin{abstract}
Some future global properties of cosmological solutions
for the Einstein-Vlasov-Maxwell system with surface symmetry are presented. Global existence is proved,
the homogeneous spacetimes are future complete for causal trajectories, and the same is true for inhomogeneous plane-symmetric solutions with small initial data. In the latter case some decay properties are also obtained at late times. Similar but slightly weaker results hold for hyperbolic symmetry. 
\end{abstract}
\section{Introduction and main results}
In general relativity the time evolution of self-gravitating collisionless particles can be modelled by the Einstein-Vlasov system. For surveys of results on that system see \cite{andreasson1}. Cosmological spacetimes are those admitting a compact Cauchy hypersurface. In this case the particles are galaxies or even clusters of galaxies.

Results on expanding cosmological models with collisionless matter, a positive cosmological constant and surface symmetry have been obtained in \cite{tchapnda1,tchapnda2}. In the present paper we want to examine what happens when the particles in the self-gravitating collisionless gas under consideration are charged. The Einstein-Vlasov system is then coupled to the Maxwell equations determining the electromagnetic field created by the charged particles. Considering this particular problem extends the knowledge on global dynamical properties of solutions of the Einstein equations. Adding the Maxwell equations could also help to answer the question why in the cosmology literature people usually talk about the magnetic field more than the electric field. If the accelerated expansion, which here is due to the positive cosmological constant, could let the latter decay faster than the former then this could be an explanation. This might also have some connection with the so-called Landau damping effect \cite{zhou}. The results of the present paper do not suffice to address this issue, but we hope they can provide a basis for doing that in the future.

As known results related to the Einstein-Vlasov-Maxwell system we can mention a small data global existence theorem in the spherically symmetric asymptotically flat setting obtained in \cite{noundjeu}. In the absence of Vlasov matter the asymptotic behaviour of solutions of that system with $T^3$-Gowdy symmetry was studied recently in \cite{ringstrom}.

Let us now formulate our system. We suppose there are two species of charged particles, one of positive charge $+1$ and the other of negative charge $-1$. All the particles are supposed to have the same rest mass equal to $1$, and to move forward in time so that the number densities $f^+$ and $f^-$ for positive and negative charge species
respectively, are non-negative functions supported on the mass
shell
\[
PM:= \{g_{\alpha\beta}p^{\alpha}p^{\beta} = -1,
 \ p^0 > 0 \},
\]
a submanifold of the tangent bundle $TM$ of the space-time
manifold $M$ with metric $g$ of signature $-+++$. We use
coordinates $(t,x^a)$ with zero shift and corresponding canonical
momenta $p^\alpha$ ; Greek indices always run from $0$ to $3$, and
Latin ones from $1$ to $3$. On the mass shell $PM$ the variable
$p^0$ becomes a function of the remaining variables $(t, x^a,
p^b)$ :
\[
p^0 = \sqrt{-g^{00}}\sqrt{1+g_{ab}p^{a}p^{b}}.
\]
The Einstein-Vlasov-Maxwell system now reads
\begin{align}\label{v1}
\partial_{t}f^+ + \frac{p^{a}}{p^{0}} \partial_{x^{a}}f^+ -
\frac{1}{p^{0}}(\Gamma_{\beta\gamma}^{a} p^{\beta}
p^{\gamma}+F_{\beta} \ ^{a}p^{\beta})
\partial_{p^{a}}f^+
 = 0
\end{align}
\begin{align}\label{v2}
\partial_{t}f^- + \frac{p^{a}}{p^{0}} \partial_{x^{a}}f^- -
\frac{1}{p^{0}}(\Gamma_{\beta\gamma}^{a} p^{\beta}
p^{\gamma}-F_{\beta} \ ^{a}p^{\beta})
\partial_{p^{a}}f^-
 = 0
\end{align}
\begin{align}\label{e}
G_{\alpha\beta} + \Lambda g_{\alpha\beta}  = 8 \pi
(T_{\alpha\beta}+\tau_{\alpha\beta})
\end{align}
\begin{align}\label{m1}
\nabla_{\alpha}F_{\beta\gamma}+\nabla_{\beta}F_{\gamma\alpha}+\nabla_{\gamma}F_{\alpha\beta}=0
\end{align}
\begin{align}\label{m2}
\nabla_{\alpha}F^{\alpha\beta}=J^{\beta}
\end{align}
\begin{align}\label{mat1}
 T_{\alpha\beta}  = - \int_{\mathbb{R}^{3}}(f^++f^-)
p_{\alpha} p_{\beta}|g|^{1/2} \frac{dp^{1}dp^{2}dp^{3}}{p_{0}}
\end{align}
\begin{align}\label{mat2}
\tau_{\alpha\beta}  = F_{\alpha\gamma}F_{\beta} \
^{\gamma}-\frac{g_{\alpha\beta}}{4}F^{\gamma\delta}F_{\gamma\delta}
\end{align}
\begin{align}\label{c}
J^{\beta} =
\int_{\mathbb{R}^{3}}(f^+-f^-)p^{\beta}|g|^{1/2}\frac{dp^{1}dp^{2}dp^{3}}{p_{0}}
\end{align}
where $p_{\alpha} = g_{\alpha\beta} p^{\beta}$,
$\Gamma_{\beta\gamma}^{\alpha}$ are the Christoffel symbols, $|g|$
denotes the determinant of the metric $g$, $G_{\alpha\beta}$ the
Einstein tensor, $\Lambda$ the cosmological constant, $F$ the
electromagnetic field created by the charged particles,
$J^{\beta}$ the total particle current density generated by the
charged particles and $T_{\alpha\beta}$ and $\tau_{\alpha\beta}$
are the energy-momentum tensor for Vlasov and Maxwell matter
respectively.

A computation in normal coordinates shows that
$\nabla_{\alpha}J^{\alpha}=0$. This equation is an expression of
the conservation of charge. It can be shown as in
\cite{rendall3} that $T_{\alpha\beta}$ satisfies the dominant
energy condition i.e. $T_{\alpha\beta}V^{\alpha}W^{\beta} \geq 0$
for any two future-pointing timelike vectors $V^{\alpha}$ and
$W^{\alpha}$. Let us show that the same is true for the Maxwell
tensor $\tau_{\alpha\beta}$. Proving this is equivalent to show
the weak-energy condition $\tau_{\alpha\beta}V^{\alpha}V^{\beta}
\geq 0$ for all timelike vector $V^{\alpha}$, together with the
property that $\tau_{\alpha\beta}V^{\beta}$ is non-spacelike for
any future-pointing timelike vector $V^{\alpha}$. The proof of the
latter can be deduced from the following identities which hold
since $F$ is antisymmetric 
\[
\tau_{\alpha\nu}\tau^{\nu}_{\beta}=\frac{1}{4}
(\tau^{\gamma\delta}\tau_{\gamma\delta})g_{\alpha\beta},
\qquad \tau_{\alpha\beta}\tau^{\alpha\beta} \geq 0
\]
Contracting the first of these identities twice with $V^{\alpha}$
implies the following, using the second identity and the fact that
$V^{\alpha}$ is timelike :
\[
(V^{\alpha}\tau_{\alpha\nu})(\tau^{\nu}_{\beta}V^{\beta})=
\frac{1}{4}(\tau^{\gamma\delta}\tau_{\gamma\delta})
g_{\alpha\beta}V^{\alpha}V^{\beta}\leq 0
\]
and setting $P_{\nu}=V^{\alpha}\tau_{\alpha\nu}$, this means that
$P_{\alpha}P^{\alpha}\leq 0$, that is $P_{\alpha}$ is
non-spacelike as desired. Now proving the weak-energy condition is
equivalent to show that $\tau_{00}$ is non-negative since we can
choose an orthonormal frame such that $V^{\alpha}$ is the timelike
vector of the frame. In such a frame $g_{00}=-1$ so that
$\tau_{00}=\frac{1}{2}g^{ab}F_{0a}F_{0b}+\frac{1}{4}F^{ab}F_{ab}
\geq 0$ as the sum of spatial lengths of a vector and a tensor
respectively.

In the present paper we adopt the definition of spacetimes with
surface symmetry, i.e., spherical, plane or hyperbolic symmetry
given in \cite{rendall1}. We write the system in areal
coordinates, i.e. coordinates are chosen such that $R=t$, where
$R$ is the area radius function on a surface of symmetry. The
circumstances under which coordinates of this type exist are
discussed in \cite{andreasson2} for the Einstein-Vlasov system
with vanishing $\Lambda$, and in \cite{tchapnda2} for the case with $\Lambda$. The analysis there
can be extended to the situation under consideration here since
the Maxwell tensor $\tau_{\alpha\beta}$ satisfies the dominant
energy condition. In such coordinates the metric takes the form
\begin{equation} \label{eq:1.1}
  ds^2 = -e^{2\mu(t,r)}dt^2 + e^{2\lambda(t,r)}dr^2 + t^2
  (d\theta^2 + \sin_{k}^{2}\theta d\varphi^{2})
\end{equation}
where
\begin{displaymath}
 \sin_{k}\theta := \left\{ \begin{array}{ll}
\sin\theta & \textrm{if $k=1$}\\
1 & \textrm{if $k=0$}\\
\sinh\theta & \textrm{if $k=-1$}
  \end{array} \right.
\end{displaymath}

Here $t > 0$, the functions $\lambda$ and $\mu$ are periodic in
$r$ with period $1$. It can be shown as in \cite{rein1} and
\cite{andreasson2} that due to the symmetry $f^+$ and $f^-$ can be
written as a function of
\[
t, r, w := e^{\lambda}p^1 \ \textrm{and} \  L := t^{4}(p^2)^2 +
t^4 \sin_{k}^{2}\theta (p^{3})^{2}, \ \textrm{with} \ r,w \in
\mathbb{R} \ ; \ L \in [0,+\infty[.
\]
In these variables we have $p^0 = e^{-\mu}\sqrt{1 + w^{2} +
L/t^{2}}$.

In surface symmetry the only non-zero components of $F$ are
$F_{01}$ and $F_{23}$. Indeed setting $h:=g+e_{0} \otimes
e_{0}-e_{1} \otimes e_{1}$, with
$e_0=e^{-\mu}\frac{\partial}{\partial t}$ and
$e_1=e^{-\lambda}\frac{\partial}{\partial r}$, the mapping
$X^\beta \mapsto h^{\beta}_{\alpha}X^\alpha$ is the orthogonal
projection on the tangent space of the orbit, and since the vector
$Y_\sigma := F_{\alpha\beta}(e_0)^{\alpha}h^{\beta}_{\sigma}$ is
invariant under the symmetry group, it vanishes. This implies that
$F_{02}=F_{03}=0$. Similarly, replacing $e_0$ by $e_1$ in the
expression of $Y_\sigma$ yields $F_{12}=F_{13}=0$.

Now we can calculate the Maxwell equations in a coordinate frame.
Equation (\ref{m1}) then implies the following, where
$(\partial_0,\partial_1,\partial_2,\partial_3)=
(\partial_t,\partial_r,\partial_\theta,\partial_\varphi)$ :
\begin{equation}\label{m3}
\partial_{0}F_{23}=\partial_{1}F_{23}=\partial_{2}F_{01}=\partial_{3}F_{01}=0.
\end{equation}
Using the fact that the mapping $(p^1, p^2, p^3)\mapsto (w, L,
p^3)$ is one-to-one from
$\mathbb{R}\times]0,\infty[\times\mathbb{R}$ to
$\mathbb{R}\times]0,\infty[\times\mathbb{R}$ and from
$\mathbb{R}\times]-\infty,0[\times\mathbb{R}$ to
$\mathbb{R}\times]0,\infty[\times\mathbb{R}$, one can compute the
$J^{\alpha}$'s and obtain $J^2=J^3=0$,
\[
J^0=-\frac{\pi}{t^2}e^{-\mu}\int_{-\infty}^{\infty}\int_{0}^{\infty}(f^+-f^-)(t,r,w,L)dLdw
\]
and
\[
J^1=-\frac{\pi}{t^2}e^{-\lambda}\int_{-\infty}^{\infty}\int_{0}^{\infty}
\frac{w}{\sqrt{1+w^2+L/t^2}}(f^+-f^-)(t,r,w,L)dLdw.
\]
Equation (\ref{m2}) then implies
\begin{equation}\label{m4}
\begin{cases}
\partial_0(\sqrt{|g|}F^{01})=J^1\sqrt{|g|}, \qquad
\partial_1(\sqrt{|g|}F^{01})=-J^0\sqrt{|g|} \\
\partial_2(\sqrt{|g|}F^{23})=0, \qquad \partial_3(\sqrt{|g|}F^{23})=0.
\end{cases}
\end{equation}
The non-zero components of the electric and magnetic parts of $F$
are $E^1=e^\mu F^{01}$ and $B_1=e^{-\mu}\sqrt{|g|}F^{23}$
respectively. Using these identities and recalling that
$\sqrt{|g|}=t^2e^{\lambda+\mu}\sin_k\theta$, equations (\ref{m3})
and (\ref{m4}) lead to the following, where $c$ is an arbitrary
constant and $E^1$ is denoted by $E$:
\begin{align*}
\partial_r(t^2e^{\lambda}E(t,r))& = \pi
e^{\lambda}\int_{-\infty}^{\infty}\int_{0}^{\infty}(f^+-f^-)(t,r,w,L)dLdw, 
\\
\partial_t(t^2e^{\lambda}E(t,r))& = -e^\mu \pi
\int_{-\infty}^{\infty}\int_{0}^{\infty}\frac{w}{\sqrt{1+w^2+L/t^2}}(f^+-f^-)(t,r,w,L)dLdw, 
\\
B_1(t,r)& = ct^{-2}e^{\lambda(t,r)}.
\end{align*}

After calculating the Vlasov equations in the variables
$(t,r,w,L)$, the non-trivial components of the Einstein tensor,
and the energy-momentum tensor and denoting by an upper dot or by
prime the derivation with respect to $t$ or $r$ respectively, the
complete Einstein-Vlasov-Maxwell system then reads as follows
\begin{equation} \label{eq:1.2}
\partial_{t}f^+ + \frac{e^{\mu-\lambda}w}{\sqrt{1+w^{2}+L/t^{2}}}
\partial_{r}f^+ - (\dot{\lambda}w +
e^{\mu-\lambda}\mu'\sqrt{1+w^{2}+L/t^{2}}-e^{\lambda+\mu}E)\partial_{w}f^+
= 0
\end{equation}
\begin{equation} \label{eq:1.3}
\partial_{t}f^- + \frac{e^{\mu-\lambda}w}{\sqrt{1+w^{2}+L/t^{2}}}
\partial_{r}f^- - (\dot{\lambda}w +
e^{\mu-\lambda}\mu'\sqrt{1+w^{2}+L/t^{2}}+e^{\lambda+\mu}E)\partial_{w}f^-
= 0
\end{equation}
\begin{equation} \label{eq:1.4}
e^{-2\mu} (2t\dot{\lambda}+1)+ k - \Lambda t^{2} = 8 \pi t^{2}\rho
\end{equation}
\begin{equation} \label{eq:1.5}
e^{-2\mu} (2t\dot{\mu}-1)- k + \Lambda t^{2} = 8 \pi t^{2}p
\end{equation}
\begin{equation} \label{eq:1.6}
\mu' = -4 \pi t e^{\lambda+\mu}j
\end{equation}
\begin{equation} \label{eq:1.7}
e^{-2\lambda}\left(\mu'' + \mu'(\mu' - \lambda')\right) -
e^{-2\mu}\left(\ddot{\lambda}+(\dot{\lambda}-
\dot{\mu})(\dot{\lambda}+\frac{1}{t})\right) + \Lambda  = 4 \pi q
\end{equation}
\begin{equation} \label{eq:1.8}
\partial_{r}(t^2 e^{\lambda}E)=t^2 e^{\lambda}a
\end{equation}
\begin{equation} \label{eq:1.9}
\partial_{t}(t^2 e^{\lambda}E)=-t^2 e^{\mu}b
\end{equation}
where
\begin{align}
\rho(t, r) & := \frac{\pi}{t^{2}} \int_{-\infty}^{\infty}
\int_{0}^{\infty} \sqrt{1+w^{2}+L/t^{2}} (f^++f^-)(t, r, w, L) dL
dw \nonumber\\
& \quad +\frac{1}{2}(e^{2\lambda}E^2+ ct^{-4})= e^{-2\mu}(T_{00}+\tau_{00})(t,
r), \label{eq:1.10}\\
p(t, r) & := \frac{\pi}{t^{2}} \int_{-\infty}^{\infty}
\int_{0}^{\infty} \frac{w^{2}}{\sqrt{1+w^{2}+L/t^{2}}}
(f^++f^-)(t, r, w, L) dL dw \nonumber\\
& \quad -\frac{1}{2}(e^{2\lambda}E^2+ ct^{-4}) =
e^{-2\lambda}(T_{11}+\tau_{11})(t, r), \label{eq:1.11}
\end{align}
\begin{align}
j(t, r) & := \frac{\pi}{t^{2}} \int_{-\infty}^{\infty}
\int_{0}^{\infty} w (f^++f^-)(t, r, w, L) dL dw = -e^{\lambda +
\mu}T_{01}(t, r), \label{eq:1.12}\\
q(t, r) & := \frac{\pi}{t^{4}} \int_{-\infty}^{\infty}
\int_{0}^{\infty} \frac{L}{\sqrt{1+w^{2}+L/t^{2}}} (f^++f^-)(t, r,
w, L) dL dw \nonumber\\  & \quad +(e^{2\lambda}E^2+ ct^{-4})=
\frac{2}{t^{2}}(T_{22}+\tau_{22})(t, r),\label{eq:1.13}\\
a(t, r) & := \frac{\pi}{t^{2}} \int_{-\infty}^{\infty}
\int_{0}^{\infty}  (f^+-f^-)(t, r, w, L) dL dw ,\label{eq:1.14}\\
b(t, r) & := \frac{\pi}{t^{2}} \int_{-\infty}^{\infty}
\int_{0}^{\infty} \frac{w}{\sqrt{1+w^{2}+L/t^{2}}} (f^+-f^-)(t, r,
w, L) dL dw.\label{eq:1.15}
\end{align}

We prescribe initial data at some time $t = t_0 > 0$,
\begin{eqnarray*}
f^+(t_0, r, w, L)= \overset{\circ}{f^+}(r, w, L), \ f^-(t_0, r, w,
L)= \overset{\circ}{f^-}(r, w, L), \\ \lambda(t_0, r) =
\overset{\circ}{\lambda}(r), \ \mu(t_0, r) =
\overset{\circ}{\mu}(r), \ E(t_0, r)=\overset{\circ}{E}(r)
\end{eqnarray*}
and want to study the existence and behaviour of the corresponding solution for
$t \in [t_0, +\infty)$.

To this end we maintain the notation in \cite{rein2,tchapnda1,tchapnda2},
and follow their work
wherever possible. The first step consists on generalizing the
local existence result in \cite[theorem 3.1]{rein2} to the case of charged particles under study:
\begin{theorem}\label{local} Let $\overset{\circ}{f^{\pm}} \in 
C^{1}(\mathbb{R}^{2} \times [0,
\infty[)$
 with $\overset{\circ}{f^{\pm}}(r+1,w,L) = 
\overset{\circ}{f^{\pm}}(r,w,L)$
 for $(r,w,L) \in \mathbb{R}^{2} \times [0, \infty[$,
 $\overset{\circ}{f^{\pm}}\geq 0$, and $w_0:=w_0^++w_0^-$,
 $L_0:=L_0^++L_0^-$ with
 \begin{align*}
 w_{0}^{\pm} &:= \sup \{ |w| | (r,w,L) \in \supp 
\overset{\circ}{f^{\pm}} \} <
 \infty\\
  L_{0}^{\pm} &:= \sup \{ L | (r,w,L) \in \supp 
\overset{\circ}{f^{\pm}} \} <
 \infty
 \end{align*}
 Let $\overset{\circ}{\lambda}, \ \overset{\circ}{E} \in 
C^{1}(\mathbb{R})$,
 $\overset{\circ}{\mu} \in C^{2}(\mathbb{R})$ with
 $\overset{\circ}{\lambda}(r) = \overset{\circ}{\lambda}(r+1)$,
$\overset{\circ}{\mu}(r) = \overset{\circ}{\mu}(r+1)$,
$\overset{\circ}{E}(r) = \overset{\circ}{E}(r+1)$ for $r \in
\mathbb{R}$, and
\begin{eqnarray*}
\overset{\circ}{\mu}'(r) =
 -4 \pi t_0 e^{\overset{\circ}{\lambda} +
\overset{\circ}{\mu}}\overset{\circ}{j}(r) = -\frac{4
\pi^{2}}{t_0}e^{\overset{\circ}{\lambda}+ \overset{\circ}{\mu}}
\int_{-\infty}^{\infty} \int_{0}^{\infty}w
(\overset{\circ}{f^+}+\overset{\circ}{f^-})(r,w,L) dL dw, \ \  r
\in \mathbb{R},
\end{eqnarray*}
\begin{eqnarray*}
\partial_r(t_0^2e^{\overset{\circ}{\lambda}}\overset{\circ}{E}) =
  t_0^2 e^{\overset{\circ}{\lambda}} \overset{\circ}{a}=
\pi e^{\overset{\circ}{\lambda}} \int_{-\infty}^{\infty}
\int_{0}^{\infty}
(\overset{\circ}{f^+}-\overset{\circ}{f^-})(r,w,L) dL dw, \ \  r
\in \mathbb{R}.
\end{eqnarray*}
Then there exists a unique, right maximal, regular solution
$(f^+,f^-, \lambda, \mu,E)$ of (\ref{eq:1.2})-(\ref{eq:1.9}) with
$(f^+,f^-, \lambda, \mu,E)(t_0) =
(\overset{\circ}{f^+},\overset{\circ}{f^-},
\overset{\circ}{\lambda},
\overset{\circ}{\mu},\overset{\circ}{E})$ on a time interval
$[t_0,T_{\max}[$ with $T_{\max} \in ]t_0,\infty]$.
\end{theorem}
The regularity concept used in this statement is defined below.

The next theorem provides a continuation criterion used to prove that 
the solution exists on the whole time interval $[t_0, \infty[$.
\begin{theorem}\label{cont}
Let $(f^+,f^-, \lambda, \mu,E)$ be a right maximal regular solution 
obtained in Theorem \ref{local}. If one has
\begin{align*}
&\sup \{ |w| | (r,w,L) \in \supp f^{+} \} <
 \infty, \ \sup \{ |w| | (r,w,L) \in \supp f^{-} \} <
 \infty, \\
&\sup \{\mu(t,r) | r \in \mathbb{R}, \ t\in[t_0,T_{max}[ \} <\infty, \ 
\sup \{(e^{\lambda}|E|)(t,r) | r \in \mathbb{R}, \ t\in[t_0,T_{max}[ \} 
<\infty
\end{align*}
 then $T_{max} =\infty$.
\end{theorem}
In the following we claim that among the conditions given in the 
previous theorem there is one which implies the others.
\begin{proposition}
The condition $$\sup \{\mu(t,r) | r \in \mathbb{R}, \ t\in[t_0,T_{max}[ 
\} <\infty$$ is sufficient in order to conclude that $T_{max}=\infty$.
\end{proposition}
We can then state
\begin{theorem}\label{global}
Consider initial data as in Theorem \ref{local} and assume in the case of spherical symmetry that $t_{0}^{2}>1/\Lambda$. Then the solution of the surface-symmetric 
Einstein-Vlasov-Maxwell system with positive cosmological constant, written in areal coordinates, exists for all $t \in[t_0, \infty[$ where $t$ denotes the area radius of the surfaces of symmetry of the induced spacetime.
\end{theorem}

Once the existence of solutions is proven, one would like to study their asymptotic behaviour at late times. In particular an important point is to know whether the spacetime obtained is future complete or not. This seems not to be easily achieved by a direct argument for the generic data case of the inhomogeneous Einstein-Vlasov-Maxwell system. Nevertheless it works out in the spatially homogeneous case, as well as in the inhomogeneous plane-symmetric case under an additional assumption on the initial data.

Let us examine spatially homogeneous solutions. These correspond to LRS (locally rotationally symmetric) models of Bianchi type I and type III and Kantowski-Sachs type for plane, hyperbolic and spherical symmetry respectively. We refer to \cite{rendall2} for a detailed discussion on these models. We use the same notation as in \cite{lee1}, where the case of uncharged particles has been studied. Most of the results obtained in that case apply also if the particles are charged, the exception being those for which the proof involves matter terms. We will also use results from \cite{wald1} in which the only requirement for the energy-momentum tensor is to satisfy the dominant and strong energy conditions, these are valid in the case under investigation. Let us formulate our system in Bianchi symmetry.

The spacetime is considered as a manifold $G \times I$, $I$ being an open interval and $G$ a simply connected three-dimensional Lie group. The metric has the form
\[
ds^2 = -d\tau^2+g_{ij}e^i \otimes e^j,
\]
where $\{e_i\}$ is a left invariant frame and $\{e^i\}$ the dual coframe.

The Einstein constraint equations are
\begin{align}
R-k_{ij}k^{ij}+(k_{ij}g^{ij})^2=16\pi (T_{00}+\tau_{00})+2\Lambda \label{c1}\\
\nabla^{i}k_{ij}=-8\pi T_{0j} \label{c2}.
\end{align}
The evolution equations are
\begin{align}
\partial_{t}g_{ij}=-2k_{ij} \label{e1} \\
\partial_{t}k_{ij}=R_{ij}+&(k_{lm}g^{lm})k_{ij}-2k_{il}k^{l}_{j}-8\pi(T_{ij} +\tau_{ij})\nonumber \\&-4\pi(T_{00}+\tau_{00})+4\pi(T_{lm}+\tau_{lm})g^{lm}g_{ij}-\Lambda g_{ij}, \label{e2}
\end{align}
where
\begin{align}
T_{00}+\tau_{00}=\int (f^++f^-)(\tau,v)&(1+g_{rs}v^rv^s)^{1/2} |g|^{1/2} dv \nonumber\\&+ F_{0\gamma}F_0 \ ^{\gamma}+\frac{1}{4}F_{\gamma \delta}F^{\gamma \delta}\\
T_{0i}=\int (f^++f^-)(\tau,v)&v_i|g|^{1/2} dv \\
T_{ij}+\tau_{ij}=\int (f^++f^-)(\tau,v)&v_iv_j(1+g_{rs}v^rv^s)^{-1/2}|g|^{1/2} dv \nonumber\\&+ F_{i\gamma}F_j \ ^{\gamma}-\frac{g_{ij}}{4}F_{\gamma \delta}F^{\gamma \delta},
\end{align}
with $v:=(v^1,v^2,v^3)$ and $dv:=dv^1dv^2dv^3$.

The Vlasov equations are
\begin{align}
\partial_\tau f^++[2k^{i}_{j}v^j-(1+g_{rs}v^rv^s)^{-1/2}\gamma_{mn}^{i}v^mv^n-(F_{0}\ ^{i}+F_{j}\ ^{i}\frac{v^j}{v^0})]\partial_{v^i}f^+=0 \label{vl1}\\
\partial_\tau f^-+[2k^{i}_{j}v^j-(1+g_{rs}v^rv^s)^{-1/2}\gamma_{mn}^{i}v^mv^n+(F_{0}\ ^{i}+F_{j}\ ^{i}\frac{v^j}{v^0})]\partial_{v^i}f^-=0,\label{vl2}
\end{align}
where $\gamma_{mn}^{i}=\frac{1}{2}g^{ik}(-C_{nk}^{l}g_{ml}+C_{km}^{l}g_{nl}+C_{mn}^{l}g_{kl})$, $C_{jk}^{i}$ are the structure constants of the Lie algebra of $G$.

Note that in LRS Bianchi symmetry the only non-zero components of $F$ are $F^{01}$ and $F^{23}$. The Maxwell equations (\ref{m1}) allow us to obtain an explicit expression for the magnetic part $F^{23}$ of $F$. Thus the remaining unknown for the Maxwell equations is the electric part $F^{01}$ of $F$. For a Bianchi model the Maxwell equations (\ref{m2}) take the following form
\[
\partial_\tau[|g|^{1/2}F^{0\beta}]+C_{ij}^iF^{j\beta}|g|^{1/2}=J^\beta|g|^{1/2}
\] 
which yields for $\beta=0$ and $\beta=i$ respectively
\begin{align}
C_{ij}^iF^{j0}=J^0 \label{c3}\\
\partial_\tau F^{0i}-(tr k)F^{0i}+C_{kj}^kF^{ji}=J^i, \label{ma3}
\end{align}
here
\begin{align*}
J^{0}&=\int (f^+-f^-)(\tau,v)|g|^{1/2} dv \\
J^{i}&=\int (f^+-f^-)(\tau,v)v^i(1+g_{rs}v^rv^s)^{-1/2}|g|^{1/2} dv.
\end{align*}
Setting $k_{ij}=\frac{1}{3}(k_{lm}g^{lm})g_{ij}+\sigma_{ij}$, $\sigma_{ij}$ being the trace free part of the second fundamental form $k_{ij}$, and using the Hamiltonian constraint (\ref{c1}) we obtain
\begin{equation*}
\frac{1}{3}(k_{ij}g^{ij})^2=\frac{-R}{2}+\frac{1}{2}\sigma_{ij}\sigma^{ij}+8\pi\rho+\Lambda,
\end{equation*}
then using the fact that $R\leq 0$ (cf. \cite{wald1}), it follows by the Proposition 1 in \cite{lee1} that the matter energy density is bounded by 
\begin{equation}\label{eq:1.37}
\rho\leq C e^{-2\gamma \tau}.
\end{equation}
We can prove the following
\begin{theorem}\label{com}
Let $f^{\pm}(0,v)$ be a nonnegative $C^1$ function with compact support. Let $(g_{ij}(0),k_{ij}(0),f^{+}(0,v),f^{-}(0,v),F^{0i}(0))$ be an initial data set for the evolution equations (\ref{e1}), (\ref{e2}), the Vlasov equations (\ref{vl1}), (\ref{vl2}) and the Maxwell equation (\ref{ma3}), which has Bianchi symmetry and satisfies the constraint equations (\ref{c1}), (\ref{c2}), and (\ref{c3}). Then the corresponding solution of the Einstein-Vlasov-Maxwell system is a future complete spacetime for causal trajectories. 
\end{theorem}
 In the inhomogeneous case the result in the latter theorem can be proved in plane symmetry for small initial data, a similar but slightly weaker result is true for hyperbolic symmetry as well. We have the following 
\begin{theorem}\label{decay}
Consider any solution of Einstein-Vlasov-Maxwell system with positive cosmological constant in plane or hyperbolic symmetry written in areal coordinates, with initial data as in theorem \ref{global}. In the case of hyperbolic symmetry assume that $\Lambda$ is sufficiently large. Let $\delta$ be a positive constant and suppose the following inequalities hold: 
\begin{equation}\label{eq:1.38}
|t_0\dot{\lambda}(t_0)-1|\leq \delta,\ |(e^{-\lambda}\mu')(t_0)|\leq \delta,\ |(e^{\lambda}E)(t_0)|\leq \delta
\end{equation}
\begin{equation}\label{eq:1.39}
|\Lambda t_{0}^2e^{2\mu(t_0)}-3-3ke^{2\mu(t_0)}|\leq \delta,\ \bar{w}(t_0)\leq \delta, \ c\leq \delta,
\end{equation}
where $\bar{w}(t)$ denotes the maximum of $|w|$ over the support of $f^+(t)$ or $f^-(t)$. Then if $\delta$ is sufficiently small, the following properties hold at late times: 
\begin{equation}\label{eq:1.40}
t\dot{\lambda}-1=O(t^{-2}),\ e^{-\lambda}\mu'=O(t^{-2}),\ e^{\lambda}E=O(t^{-2}),
\end{equation}
\begin{equation}\label{eq:1.41}
\Lambda t^2e^{2\mu}-3-3ke^{2\mu}=O(t^{-3}),\ \bar{w}=O(t^{-1}).
\end{equation}
Furthermore the spacetime is future complete for causal trajectories. 
\end{theorem}
The rest of the paper is organized as follows. In section 2 we
present some preliminary results that we use to prove Theorem
\ref{local} in section 3. The proof for the other results is also given 
in section 3.
\section{Preliminaries}
The regularity properties required for a solution are as in
\cite{rein2}.
\begin{definition} \label{d1} Let $I \subset ]0, \infty[$ be
an interval \\
(a) $f^{\pm} \in C^{1}(I \times \mathbb{R}^{2} \times [0,
\infty[)$ is regular, if $f^{\pm}(t, r+1, w, L) = f^{\pm}(t, r, w,
L)$ for $(t, r, w, L) \in I \times \mathbb{R}^{2} \times [0,
\infty[$, $f^{\pm} \geq 0$, and ${\rm supp}f^{\pm}(t, r, ., .)$ is
compact, uniformly in $r$ and locally
uniformly in $t$.\\
(b) $\rho$ (or $p$, $j$, $q$, $a$, $b$)$ \in C^{1}(I \times
\mathbb{R})$ is regular, if $\rho(t, r+1) = \rho(t, r)$ for $(t,
r) \in I \times
\mathbb{R}$ \\
(c) $\lambda \in C^{1}(I \times \mathbb{R})$ is regular, if
$\dot{\lambda} \in C^{1}(I \times \mathbb{R})$ and $\lambda(t,
r+1) = \lambda(t, r)$ for $(t, r) \in I \times
\mathbb{R}$ \\
(d) $\mu\in C^{1}(I \times \mathbb{R})$ is regular, if $\mu' \in
C^{1}(I \times \mathbb{R})$ and $\mu(t, r+1) = \mu(t, r)$ for $(t,
r) \in I \times \mathbb{R}$.\\
(e) $E$ (or $\tilde{\mu}$) $\in C^{1}(I \times \mathbb{R})$ is
regular, if $E(t, r+1) = E(t, r)$ for $(t, r) \in I \times
\mathbb{R}$.
\end{definition}
It is possible to solve each equation of the system
(\ref{eq:1.2})-(\ref{eq:1.5}), (\ref{eq:1.9}), when the other
unknowns are given. This is the content of the following
\begin{proposition} \label{p:2.2}
Let $\bar{f^+}$, $\bar{f^-}$, $\bar{\lambda}$, $\bar{\mu}$,
$\bar{E}$ be regular for $(t,r)\in I\times\mathbb{R}$, $I \subset
]0, \infty[$ an interval with $t_0 \in I$. Replace $f^+$, $f^-$,
$\lambda$, $\mu$, $E$ respectively by $\bar{f^+}$, $\bar{f^-}$,
$\bar{\lambda}$, $\bar{\mu}$, $\bar{E}$ in $\rho$, $b$, $p$ to
define $\bar{\rho}$, $\bar{b}$ and $\bar{p}$. Suppose that
$\overset{\circ}{f^+}, \ \overset{\circ}{f^-} \in
C^{1}(\mathbb{R}^2\times[0,\infty[)$, $\overset{\circ}{\lambda}, \
\overset{\circ}{\mu}, \ \overset{\circ}{E} \in C^{1}(\mathbb{R})$
and are periodic of period $1$ in $r$. Assume that
\begin{equation} \label{eq:2.1}
\frac{t_{0}(e^{-2 \overset{\circ}{\mu}(r)} + k)}{t} - k - \frac{8
\pi}{t}\int_{t_0}^{t}s^{2}\bar{p}(s, r) ds +
\frac{\Lambda}{3t}(t^{3} - t_{0}^{3} ) > 0 , (t,r) \in I \times
\mathbb{R}.
\end{equation}
Then the system
\begin{equation} \label{eq:2.2}
\partial_{t}f^+ + 
\frac{e^{\bar{\mu}-\bar{\lambda}}w}{\sqrt{1+w^{2}+L/t^{2}}}
\partial_{r}f^+ - (\dot{\bar{\lambda}}w +
e^{\bar{\mu}-\bar{\lambda}}\bar{\mu}'\sqrt{1+w^{2}+L/t^{2}}-e^{\bar{\lambda}+\bar{\mu}}\bar{E})\partial_{w}f^+
= 0
\end{equation}
\begin{equation} \label{eq:2.3}
\partial_{t}f^- + 
\frac{e^{\bar{\mu}-\bar{\lambda}}w}{\sqrt{1+w^{2}+L/t^{2}}}
\partial_{r}f^- - (\dot{\bar{\lambda}}w +
e^{\bar{\mu}-\bar{\lambda}}\bar{\mu}'\sqrt{1+w^{2}+L/t^{2}}+e^{\bar{\lambda}+\bar{\mu}}\bar{E})\partial_{w}f^-
= 0
\end{equation}
\begin{equation} \label{eq:2.4}
e^{-2\mu} (2t\dot{\lambda}+1)+ k - \Lambda t^{2} = 8 \pi
t^{2}\bar{\rho}
\end{equation}
\begin{equation} \label{eq:2.5}
e^{-2\mu} (2t\dot{\mu}-1)- k + \Lambda t^{2} = 8 \pi t^{2}\bar{p}
\end{equation}
\begin{equation} \label{eq:2.6}
\partial_{t}(t^2 e^{\lambda}E)=-t^2 e^{\mu}\bar{b}
\end{equation}
has a unique, regular solution $(f^+, f^-, \lambda, \mu, E)$ on $I
\times \mathbb{R}$ with $f^+(t_0) = \overset{\circ}{f^+}$,
$f^-(t_0) = \overset{\circ}{f^-}$, $\lambda(t_0) =
\overset{\circ}{\lambda}$, $\mu(t_0) = \overset{\circ}{\mu}$ and
$E(t_0) = \overset{\circ}{E}$. The solution is given by
\begin{equation}\label{eq:2.7}
  f^{\pm}(t,r,w,L) = \overset{\circ}{f^{\pm}}((R^{\pm},W^{\pm})(t_0, t, 
w,L), L)
\end{equation}
\begin{equation} \label{eq:2.8}
e^{-2 \mu(t, r)} = \frac{t_{0}(e^{-2 \overset{\circ}{\mu}(r)} +
k)}{t} - k - \frac{8 \pi}{t}\int_{t_0}^{t}s^{2}\bar{p}(s, r) ds
+ \frac{\Lambda}{3t}(t^{3} - t_{0}^{3} )
\end{equation}
\begin{equation} \label{eq:2.9}
\dot{\lambda}(t, r) = 4 \pi t e^{2 \mu(t, r)}\bar{\rho}(t, r) -
\frac{1 + k e^{2 \mu(t, r)} }{2t} + \frac{\Lambda}{2}t e^{2 \mu(t,
r)}
\end{equation}
\begin{equation} \label{eq:2.10}
\lambda(t, r) = \overset{\circ}{\lambda}(r) +
\int_{t_0}^{t}\dot{\lambda}(s, r) ds
\end{equation}
\begin{equation}\label{eq:2.11}
E(t,r)=t^{-2}e^{-\lambda(t,r)}
\left(t_0^2e^{\overset{\circ}{\lambda}(r)}\overset{\circ}{E}(r)-
\int_{t_0}^{t}s^2e^{\mu(s,r)}\bar{b}(s,r)ds\right)
\end{equation}
where $(R^{\pm},W^{\pm})$ is the solution of the characteristic
system
\begin{equation}\label{eq:2.12}
\frac{d}{ds}(r,w)=(\frac{e^{\bar{\mu}-\bar{\lambda}}w}{\sqrt{1+w^{2}+L/t^{2}}},
-\dot{\bar{\lambda}}w -
e^{\bar{\mu}-\bar{\lambda}}\bar{\mu}'\sqrt{1+w^{2}+L/t^{2}}\pm
e^{\bar{\lambda}+\bar{\mu}}\bar{E})
\end{equation}
satisfying $(R^{\pm},W^{\pm})(t,t,r,w,L)=(r,w)$.
 If $I = [t_0, T[$ with $T \in ]t_0, \infty]$ then there exists some
$T^{\star} \in ]t_0, T]$ such that condition (\ref{eq:2.1}) holds
on $[t_0, T^{\star}[ \times \mathbb{R}$. $T^{\star}$ depends on
$\bar{p}$.
\end{proposition}
{\bf Proof} Integrating (\ref{eq:2.6}) with respect to $t$ over
$[t_0,t]$ gives (\ref{eq:2.11}). The rest of the proof is similar
to those of propositions 2.2 and 2.3, 1) in \cite{tchapnda1}.
$\Box$

In order to solve the system (\ref{eq:1.2})-(\ref{eq:1.9}), it
will be enough to concentrate on the subsystem
(\ref{eq:1.2})-(\ref{eq:1.5}), (\ref{eq:1.9}), as can be seen in
\begin{proposition}\label{p:2.3}
The subsystem (\ref{eq:1.2})-(\ref{eq:1.5}), (\ref{eq:1.9}) is
equivalent to the full system (\ref{eq:1.2})-(\ref{eq:1.9}),
provided the initial data satisfy (\ref{eq:1.6}) and
(\ref{eq:1.8}) at $t=t_0$.
\end{proposition}
{\bf Proof} Under the assumption that the subsystem
(\ref{eq:1.2})-(\ref{eq:1.5}), (\ref{eq:1.9}) is satisfied as well
as equations (\ref{eq:1.6}) and (\ref{eq:1.8}) for $t=t_0$, we
should prove that (\ref{eq:1.6})-(\ref{eq:1.8}) hold for all $t$.
Integrating (\ref{eq:1.9}) over $[t_0,t]$ with respect to t and
differentiating the resulting equation with respect to $r$ yields
\begin{equation}\label{eq:2.13}
\partial_{r}(t^2e^\lambda
E)=t_0^2e^{\overset{\circ}{\lambda}}\overset{\circ}{a}+\int_{t}^{t_0}s^2e^\mu(b'+\mu'b)ds.
\end{equation}
Using equations (\ref{eq:1.2}) and (\ref{eq:1.3}) and integration
by parts with respect to $t$ and $w$ leads to an expression for
$\int_{t}^{t_0}s^2e^\mu b'ds$ so that (\ref{eq:2.13}) implies
\begin{equation}\label{eq:2.14}
\partial_{r}(t^2e^\lambda
E)=t^2e^{\lambda}a+\pi\int_{t}^{t_0}\int_{-\infty}^{\infty}
\int_{0}^{\infty}e^{2\lambda+\mu}E(f^++f^-)(s,r,w,L)dLdwds.
\end{equation}
Computing the conservation law $\nabla_\alpha J^\alpha=0$ in
coordinates $(t,r,w,L)$ and integrating the resulting equation
with respect to $t$ yields
\begin{equation*}
\pi\int_{t}^{t_0}\int_{-\infty}^{\infty}
\int_{0}^{\infty}e^{2\lambda+\mu}E(f^++f^-)(s,r,w,L)dLdwds=0
\end{equation*}
and so (\ref{eq:2.14}) becomes $\partial_{r}(t^2e^\lambda
E)=t^2e^{\lambda}a$ that is (\ref{eq:1.8}) holds for all $t$.
Using the latter and an argument similar to the one used in the proof for \cite[Prop.
2.2]{rein2} we can show that (\ref{eq:1.6}) and (\ref{eq:1.7})
hold for all $t$ as well. $\Box$

The latter proposition shows that equations (\ref{eq:1.6}) and
(\ref{eq:1.8}) are invariant under evolution. So they will be
considered as constraint equations on initial data (at $t=t_0$).
They can be solved :
\begin{proposition}\label{p:2.4}
The constraint equations $\overset{\circ}{\mu}'(r) =
 -4 \pi t_0 e^{\overset{\circ}{\lambda} +
\overset{\circ}{\mu}}\overset{\circ}{j}(r)$ and\\
$\partial_r(t_0^2e^{\overset{\circ}{\lambda}}\overset{\circ}{E}) =
  t_0^2 e^{\overset{\circ}{\lambda}} \overset{\circ}{a}$ are
  solvable.
\end{proposition}
{\bf Proof} To solve these equations we need to impose the
following conditions, because
$(e^{\overset{\circ}{\lambda}}\overset{\circ}{E})(r)$ and 
$e^{-\overset{\circ}{\mu}(r)}$ are periodic
in $r$ with period $1$ :
\[
\int_{0}^{1}\int_{-\infty}^{\infty} \int_{0}^{\infty}
e^{\overset{\circ}{\lambda}}(\overset{\circ}{f^+}-\overset{\circ}{f^-})(r,w,L)
dL dw dr=0,
\]
and
\begin{eqnarray*}
I(\overset{\circ}{f^{\pm}}) := \frac{4 \pi^{2}}{t_0}\int_{0}^{1}
\int_{-\infty}^{\infty} \int_{0}^{\infty}
e^{\overset{\circ}{\lambda}}w \overset{\circ}{f^{\pm}}(r,w,L) dL dw dr 
=0
\end{eqnarray*}
Choosing $\overset{\circ}{\lambda}$ freely, the argument of the proof in
\cite[Remark 2.4]{tchapnda2} applies. $\Box$
\\
\\
{\bf Remark} Note that considering a model with more than one
species of particles is important in order to prove the
solvability of the second constraint equation above. Indeed if we
had only one species of particles the integral
$\int_{0}^{1}\int_{-\infty}^{\infty} \int_{0}^{\infty}
e^{\overset{\circ}{\lambda}}\overset{\circ}{f}(r,w,L) dL dw dr$
would never vanish, except if $\overset{\circ}{f}$ is identically zero.
\section{Proofs}
\subsection{Proof of Theorem \ref{local}}
Instead of considering the subsystem
(\ref{eq:1.2})-(\ref{eq:1.5}), (\ref{eq:1.9}), an idea used in
\cite{rein2}, that we follow here, is to consider an auxiliary
system consisting of
\begin{equation} \label{eq:3.1}
\partial_{t}f^+ + \frac{e^{\mu-\lambda}w}{\sqrt{1+w^{2}+L/t^{2}}}
\partial_{r}f^+ - (\dot{\lambda}w +
e^{\mu-\lambda}\tilde{\mu}\sqrt{1+w^{2}+L/t^{2}}-e^{\lambda+\mu}E)\partial_{w}f^+
= 0,
\end{equation}
\begin{equation} \label{eq:3.2}
\partial_{t}f^- + \frac{e^{\mu-\lambda}w}{\sqrt{1+w^{2}+L/t^{2}}}
\partial_{r}f^- - (\dot{\lambda}w +
e^{\mu-\lambda}\tilde{\mu}\sqrt{1+w^{2}+L/t^{2}}+e^{\lambda+
\mu}E)\partial_{w}f^- = 0,
\end{equation}
together with (\ref{eq:1.4}), (\ref{eq:1.5}), (\ref{eq:1.9}) and
\begin{equation} \label{eq:3.3}
\tilde{\mu} = -4 \pi t e^{\lambda+\mu}j.
\end{equation}
Next by proving that $\mu'=\tilde{\mu}$ it is easy to show that if $(f^+,f^-,\lambda,\mu,\tilde{\mu},E)$ is
a regular solution  of 
(\ref{eq:3.1}), (\ref{eq:3.2}), (\ref{eq:1.4}), (\ref{eq:1.5}), 
(\ref{eq:3.3}), 
(\ref{eq:1.9}) on some time interval $I \subset ]0,\infty[$ with 
$t_0\in 
I$, and with initial data satisfying (\ref{eq:1.6}) and (\ref{eq:1.8}) 
for $t=t_0$, then $(f^+,f^-,\lambda,\mu,E)$ solves 
(\ref{eq:1.2})-(\ref{eq:1.9}).

As in \cite{rein2}, the solution of the auxiliary system above is
used to construct a sequence of iterative solutions.

Let $\overset{\circ}{\tilde{\mu}} := \overset{\circ}{\mu}'$,
$\lambda_{0}(t,r) := \overset{\circ}{\lambda}(r)$, $\mu_{0}(t,r)
:= \overset{\circ}{\mu}(r)$, $\tilde{\mu}_{0}(t,r) :=
\overset{\circ}{\tilde{\mu}}(r)$, $E_{0}(t,r) :=
\overset{\circ}{E}(r)$ for $t \in [t_0, \infty[$, $r \in \mathbb{R}$ ;
$T_{0} = \infty$. If $\lambda_{n-1}$, $\mu_{n-1}$,
$\tilde{\mu}_{n-1}$, $E_{n-1}$ are already defined and regular on
$[t_0, T_{n-1}[ \times \mathbb{R}$ with $T_{n-1} \geq 0$ then let
\begin{align} \label{eq:3.4}
  G_{n-1}^{\pm}(t, r, w, L)& :=
  (\frac{w e^{\mu_{n-1}-\lambda_{n-1}}}{\sqrt{1+w^{2}+L/t^{2}}},
  -\dot{\lambda}_{n-1}w  \nonumber\\
   &- e^{\mu_{n-1}-\lambda_{n-1}}\tilde{\mu}_{n-1}
  \sqrt{1+w^{2}+L/t^{2}}\pm e^{\lambda_{n-1}+\mu_{n-1}}E_{n-1})
\end{align}
for $t \in [t_0, T_{n-1}[$ and denote by $(R_{n}^{\pm},
W_{n}^{\pm})(s, t, r, w, L)$ the solution of the characteristic
system
\begin{eqnarray*}
\frac{d}{ds}(R^{\pm}, W^{\pm}) = G_{n-1}^{\pm}(s, R, W, L)
\end{eqnarray*}
with initial data
\begin{eqnarray*}
(R_{n}^{\pm}, W_{n}^{\pm})(t,t,r,w,L) = (r, w), \ \ (t, r, w,
L)\in [t_0, T_{n-1}[ \times \mathbb{R}^{2} \times [0, \infty[.
\end{eqnarray*}
Define
\begin{eqnarray*}
f_{n}^{\pm}(t, r, w, L) :=
\overset{\circ}{f^{\pm}}\left((R_{n}^{\pm}, W_{n}^{\pm})(t_0, t,
r, w, L), L\right),
\end{eqnarray*}
that is, $f_{n}^{\pm}$ is the solution of
\begin{align} \label{eq:3.5}
\partial_{t}f_{n}^{\pm}& + \frac{w
e^{\mu_{n-1}-\lambda_{n-1}}}{\sqrt{1+w^{2}+L/t^{2}}}\partial_{r}f_{n}^{\pm}
- (\dot{\lambda}_{n-1}w  +
\nonumber\\
&
e^{\mu_{n-1}-\lambda_{n-1}}\tilde{\mu}_{n-1}\sqrt{1+w^{2}+L/t^{2}}\mp
e^{\lambda_{n-1}+\mu_{n-1}}E_{n-1})\partial_{w}f_{n}^{\pm} = 0
\end{align}
with $f_{n}^{\pm}(t_0) = \overset{\circ}{f^{\pm}}$, and define
$\rho_{n}$, $p_{n}$, $j_{n}$, $q_{n}$, $a_{n}$, $b_{n}$ by the
integrals (\ref{eq:1.10})-(\ref{eq:1.15}) with $f^{\pm}$, $E$,
$\lambda$ replaced by $f_{n}^{\pm}$, $E_{n-1}$, $\lambda_{n-1}$
respectively. Define
\begin{eqnarray*}
 T_{n}  :=  \sup \left\{ t' \in ]t_{0}, T_{n-1}[ \ |
\frac{t_{0}(e^{-2 \overset{\circ}{\mu}(r) } + k)}{t} - k - \frac{8
\pi}{t}\int_{t_{0}}^{t}s^{2}p_{n}(s, r) ds\right.{}\\
\left. {}
 + \frac{\Lambda}{3t}(t^{3} - t_{0}^{3}) > 0, r \in \mathbb{R}, t
\in [t_{0}, t']\right\}.
\end{eqnarray*}
Using Proposition \ref{p:2.2}, let
\begin{equation} \label{eq:3.6}
e^{-2 \mu_n(t, r)} := \frac{t_{0}(e^{-2 \overset{\circ}{\mu}(r)} +
k)}{t} - k - \frac{8 \pi}{t}\int_{t_0}^{t}s^{2}p_n(s, r) ds +
\frac{\Lambda}{3t}(t^{3} - t_{0}^{3} ),
\end{equation}
\begin{equation} \label{eq:3.7}
\dot{\lambda}_n(t, r) := 4 \pi t e^{2 \mu_n(t, r)}\rho_n(t, r) -
\frac{1 + k e^{2 \mu_n(t, r)} }{2t} + \frac{\Lambda}{2}t e^{2
\mu_n(t, r)},
\end{equation}
\begin{equation} \label{eq:3.8}
\lambda_n(t, r) := \overset{\circ}{\lambda}(r) +
\int_{t_0}^{t}\dot{\lambda}_n(s, r) ds,
\end{equation}
\begin{equation} \label{eq:3.9}
\tilde{\mu}_n(t,r) := -4 \pi t e^{\lambda_n+\mu_n}j_n(t,r),
\end{equation}
\begin{equation}\label{eq:3.10}
E_n(t,r):=t^{-2}e^{-\lambda_n(t,r)}
\left(t_0^2e^{\overset{\circ}{\lambda}(r)}\overset{\circ}{E}(r)-
\int_{t_0}^{t}s^2e^{\mu_n(s,r)}b_n(s,r)ds\right).
\end{equation}
Subtracting equation (\ref{eq:3.5}) corresponding to $f^+$ by the
one corresponding to $f^-$, integrating the resulting equation
with respect to $w$ and $L$ and integrating by parts with respect
to $w$ yield
\begin{equation*}
\partial_{r}(t^2e^{\mu_n}b_n)=(\dot{\lambda}_n-\dot{\lambda}_{n-1})
e^{\lambda_{n-1}-\mu_{n-1}+\mu_{n}}t^2a_n-e^{\lambda_{n-1}-\lambda_{n}
-\mu_{n-1}+\mu_{n}}\partial_{t}(t^2e^{\lambda_n}a_n),
\end{equation*}
so that multiplying (\ref{eq:3.10}) by $t^2e^{\lambda_n}$ and
differentiating the resulting equation with respect to $r$ lead to
\begin{align}\label{eq:3.11}
\partial_{r}(t^2e^{\lambda_n}E_n)=& 
t^2a_ne^{\lambda_{n-1}-\mu_{n-1}}-\int_{t_0}^{t}s^2
(\mu_{n}'-\tilde{\mu}_{n-1})e^{\mu_{n}}b_n\ ds \nonumber
\\ & +\int_{t_0}^{t}s^2(\dot{\mu}_{n}-\dot{\mu}_{n-1})
a_ne^{\mu_{n}-\mu_{n-1}+\lambda_{n-1}}\ ds.
\end{align}
We split the proof of theorem \ref{local} into several lemmas.
From now on $\| \cdot \|$ denotes the $L^\infty$-norm on the
function space in question, the numerical constant $C$ may change from line to line and does not depend on $n$ or $t$ or the initial
data. Firstly we prove:
\begin{lemma}\label{l1}
The sequences $\mu_n$, $\lambda_n$, $\dot{\lambda}_n$, $\rho_n$,
$p_n$, $E_n$, $j_n$, $a_n$, $b_n$, $\tilde{\mu}_n
e^{\mu_n-\lambda_n}$ are uniformly bounded in $n$, in the $L^{\infty}$-norm by a continuous function on $[t_0,\infty[$.
\end{lemma}
{\bf Proof} Define $P_n(t):=(P^+_n+P^-_n)(t)$ with
\begin{align*} P_{n}^{\pm}(t) &:= \sup \{ |w||(r,w,L) \in {\rm
supp}f_{n}^{\pm}(t) \}, \quad t\in[t_0,T_n[, \quad {\rm and}\\
Q_{n}(t) &:= \sup \{ se^{2\mu_n(s,r)}|r \in \mathbb{R}, t_0\leq
s\leq t \},\\
S_{n}(t) &:= \sup \{ |E_n|e^{\lambda_n(s,r)}|r \in
\mathbb{R}, t_0\leq s\leq t \},
\end{align*}
we have the following estimate on $\supp f_{n}^{\pm}(t)$
\begin{eqnarray*}
\sqrt{1 + w^{2} + L/t^{2}} \leq \sqrt{1 + (P_{n}^{\pm}(t))^{2} +
L_{0}^{\pm}/t_{0}^{2}} \leq C(1 + L_{0}^{\pm})(1 +
P_{n}^{\pm}(t)),
\end{eqnarray*}
so that
\begin{align}
\parallel \rho_{n}^{\pm}(t)\parallel, \parallel p_{n}^{\pm}(t)\parallel 
&\leq \frac{C}{t^2} (1 + L_{0}^{\pm})^{2}
(1 + \parallel \overset{\circ}{f^{\pm}} \parallel)(1 +
P_{n}^{\pm}(t))^{2}(1+S_{n-1}(t))^2 \label{eq:3.12}\\
\parallel j_{n}^{\pm}(t)\parallel &\leq \frac{C}{t} (1 + 
L_{0}^{\pm})^{2}
(1 + \parallel \overset{\circ}{f^{\pm}} \parallel)(1 +
P_{n}^{\pm}(t))^{2}, \label{eq:3.12}
\end{align}
and then using (\ref{eq:3.7}) and (\ref{eq:3.9})
\begin{align} \label{eq:3.13}
    \mid e^{\mu_{n}-\lambda_{n}} \tilde{\mu}_{n}(t,r)\mid + \mid
\dot{\lambda}_{n}(t,r)\mid \leq C(1+\Lambda)&(1+L_{0}^{\pm})^{2}
 (1+\parallel
 \overset{\circ}{f^{\pm}}\parallel)(1+P_{n}^{\pm}(t))^{2}\nonumber\\
 &(1+Q_{n}(t))(1+S_{n-1}(t))^2
\end{align}
This inequality is used to obtain an estimate on $\supp
f_{n+1}^{\pm}(t)$ for $|\dot{W}_{n+1}^{\pm}|$ which implies
\begin{align*}
    P^{\pm}_{n+1}(t) \leq w_{0}^{\pm}+C^*\int_{t_0}^{t}&
 (1+P_{n}^{\pm}(s))^{2}(1+P_{n+1}^{\pm}(s))
 (1+Q_{n}(s))^{3/2}(1+S_{n-1}(s))^2\\&(1+S_{n}(s))\ ds,
\end{align*}
with $C^*=C(1+\Lambda)(1+L_{0}^{\pm})^{2}(1+\parallel
 \overset{\circ}{f^{\pm}}\parallel)$. Setting
 $\tilde{P}_{n}^{\pm}(t):=\sup\{P^{\pm}_{m}(t)|m \leq n\}$ and $\tilde{S}_{n}(t):=\sup\{S_{m}(t)|m \leq n\}$, it
 follows that
\begin{equation*}
    \tilde{P}^{\pm}_{n+1}(t) \leq w_{0}^{\pm}+C^*\int_{t_0}^{t}
 (1+\tilde{P}_{n+1}^{\pm}(s))^{3}
 (1+Q_{n}(s))^{3/2}(1+\tilde{S}_{n}(s))^3 ds,
\end{equation*}
whence
\begin{equation}\label{eq:3.14}
    \tilde{P}_{n+1}(t) \leq w_{0}+C^*\int_{t_0}^{t}
 (1+\tilde{P}_{n+1}(s))^{3}
 (1+Q_{n}(s))^{3/2}(1+\tilde{S}_{n}(s))^3 ds.
\end{equation}
Taking the derivative of (\ref{eq:3.6}) with respect to $t$ leads
to
\begin{eqnarray*}
 t(2\dot{\mu}_{n}e^{2\mu_n}) = 8\pi (t e^{2\mu_{n}})^2 p_{n} +
 \frac{k}{t^2}(te^{2\mu_{n}})^2+\frac{1}{t}te^{2\mu_{n}}- \Lambda(t
 e^{2\mu_{n}})^2,
\end{eqnarray*}
integrating this over $[t_0,t]$ and using integration by parts for
the left hand side yields
\begin{equation}\label{eq:3.15}
    Q_{n}(t) \leq \|t_{0}e^{2\overset{\circ}{\mu}}\|+C^*\int_{t_0}^{t}
 (1+\tilde{P}_{n}(s))
 (1+Q_{n}(s))^{2}(1+\tilde{S}_{n-1}(s))^2 ds.
\end{equation}
Next (\ref{eq:3.10}) implies that
\begin{equation}\label{eq:3.16}
    S_{n}(t) \leq 
\|e^{\overset{\circ}{\lambda}}\overset{\circ}{E}\|+C^*\int_{t_0}^{t}
 (1+\tilde{P}_{n}(s))^2
 (1+Q_{n}(s))^{1/2} ds.
\end{equation}
Adding (\ref{eq:3.14}), (\ref{eq:3.15}) and (\ref{eq:3.16})
implies that
\[
(\tilde{P}_{n+1}+Q_n+\tilde{S}_{n})(t) \leq 
w_0+\|t_{0}e^{2\overset{\circ}{\mu}}\|+    
\|e^{\overset{\circ}{\lambda}}\overset{\circ}{E}\|+C^{*}\int_{t_0}^{t}
[1+(\tilde{P}_{n+1}+Q_n+\tilde{S}_{n})(s)]^{15/2} ds.
\]
Thus $P_n$, $Q_n$, $S_n$ are estimated by $z_1(t)$, the right
maximal solution of the equation
\[
    z_1(t) = w_0+\|t_{0}e^{2\overset{\circ}{\mu}}\|+    
\|e^{\overset{\circ}{\lambda}}\overset{\circ}{E}\|+C^*\int_{t_0}^{t}
 (1+z_1(s))^{15/2} ds,
\]
which exists on $[t_0, T^{(1)}[$, we have $T_n \geq T^{(1)}$.
Therefore there exists a continuous function $C_1(t)$ which
depends only on $z_1$ as an increasing function such that $\mu_n$,
$\lambda_n$, $\dot{\lambda}_n$, $\rho_n$, $p_n$, $E_n$, $j_n$,
$a_n$, $b_n$, $\tilde{\mu}_n e^{\mu_n-\lambda_n}$ are bounded in
the $L^\infty$-norm by $C_1(t)$. $\Box$

Next we prove :
\begin{lemma}\label{l2}
The sequences $\mu'_n$, $\lambda'_n$, $\dot{\lambda}'_n$,
$\rho'_n$, $p'_n$, $E'_n$, $j'_n$, $a'_n$, $b'_n$, $\tilde{\mu}'_n
$ are uniformly bounded in $n$, in the $L^\infty$-norm by a continuous function on $[t_0,\infty[$.
\end{lemma}
\textbf{Proof} Let us start from equation (\ref{eq:3.11}). Given
Lemma \ref{l1} most of the quantities on the right hand side of
(\ref{eq:3.11}) can be estimated. The exception is $\mu'_n$ which
is obtained after differentiating (\ref{eq:3.6}) with respect to
$r$ :
\begin{eqnarray*}
\mu_{n}'(t, r) = \frac{e^{2\mu_{n}}}{t}\left(t_{0}
\overset{\circ}{\mu}'(r) e^{-2 \overset{\circ}{\mu}} - 4 \pi
\int_{t}^{t_{0}}s^{2} p_{n}'(s, r)ds\right),
\end{eqnarray*}
again most of the terms are unproblematic with one exception that
is $p'_n$ :
\begin{align*}
 p'_n(t, r) & = \frac{\pi}{t^{2}} \int_{-\infty}^{\infty}
\int_{0}^{\infty} \frac{w^{2}}{\sqrt{1+w^{2}+L/t^{2}}}
\partial_r(f_{n}^{+}+f_{n}^{-})(t, r, w, L) dL
dw\\
& \
-e^{\lambda_{n-1}}E_{n-1}\partial_{r}(e^{\lambda_{n-1}}E_{n-1}).
\end{align*}
Defining
\begin{eqnarray*}
D_{n}(t) := \sup \{ \parallel \partial_{r}f_{n}^{+}(s) \parallel +
\parallel \partial_{r}f_{n}^{-}(s) \parallel | t_0 \leq s \leq t
\},
\end{eqnarray*}
and
\begin{eqnarray*}
\Delta_{n}(t) := \sup \{ \parallel \partial_{r}(e^{\lambda_n}E_n)(s)
\parallel | t_0 \leq s \leq t \},
\end{eqnarray*}
we deduce the following estimates
\begin{align*}
 \parallel t^2 p_{n}'(t)
\parallel & \leq C_{1}(t)(D_n(t)+\Delta_{n-1}(t)) \\
\parallel \mu_{n}'(t)
\parallel & \leq C_{1}(t)(c_1+D_n(t)+\Delta_{n-1}(t)),
\end{align*}
where $c_{1} := \parallel e^{-2
\overset{\circ}{\mu}}\overset{\circ}{\mu}' \parallel +
\parallel \overset{\circ}{\lambda}' \parallel + 1 + \mid \Lambda
\mid$, and so equation (\ref{eq:3.11}) implies
\begin{equation*}
\parallel \partial_{r}(e^{\lambda_n}E_n)(t)
\parallel \leq
C_1(t)+\int_{t_{0}}^{t}C_{1}(s)\left(c_1+D_n(s)+\Delta_{n-1}(s)\right)ds
\end{equation*}
thus
\begin{equation}\label{eq:3.18}
\Delta_n(t) \leq
C_1(t)+\int_{t_{0}}^{t}C_{1}(s)\left(c_1+D_n(s)+\Delta_{n-1}(s)\right)ds.
\end{equation}
Now differentiating (\ref{eq:3.7}) and (\ref{eq:3.8}) yields
\begin{align*}
\dot{\lambda}_{n}'(t, r) & = e^{2\mu_{n}}\left(8 \pi t \mu_{n}'(t,
r) \rho_{n}(t, r) + 4 \pi t \rho_{n}'(t, r) - \frac{k}{t}
\mu_{n}'(t, r) + \Lambda t \mu_{n}'\right) \\
\lambda_{n}'(t, r) & = \overset{\circ}{\lambda}'(r) +
\int_{t_{0}}^{t} \dot{\lambda}_{n}'(s, r) ds,
\end{align*}
and using the expression for $\rho'_n$ we obtain
\begin{align*}
 \parallel t \rho_{n}'(t)
\parallel & \leq C_{1}(t)(D_n(t)+\Delta_{n-1}(t)) \\
\parallel \lambda_{n}'(t)
\parallel, \ \parallel \dot{\lambda}_{n}'(t)
\parallel & \leq C_{1}(t)(c_1+D_n(t)+\Delta_{n-1}(t)).
\end{align*}
On the other hand it follows from (\ref{eq:3.9}) that
$$
e^{\mu_{n} - \lambda_{n}} \tilde{\mu}_{n} = -4 \pi t e^{2 \mu_{n}}
j_{n},
$$ and
$$
\mid (e^{\mu_{n} - \lambda_{n}} \tilde{\mu}_{n})'(t, r)\mid \leq
C_{1}(t) (c_{1} + D_{n}(t)+\Delta_{n-1}(t)).
$$
We can now estimate the derivatives of $G_{n}^{\pm}$ with respect
to $r$ and $w$ :
\begin{align*}
\lefteqn{\partial_{r}G_{n}^{\pm}(t, r, w, L) = ((\mu_{n} -
\lambda_{n})'e^{\mu_{n}- \lambda_{n}}\frac{w}{\sqrt{1+w^{2}+L/t^{2}}}, 
} \\
& & {} & & {} & & {} & & {} & & {} -(e^{\mu_{n} - \lambda_{n}}
\tilde{\mu}_{n})'\sqrt{1+w^{2}+L/t^{2}} - \dot{\lambda}_{n}'
w\pm(\mu'_ne^{\mu_n+\lambda_n}E_n+e^{\mu_n}\partial_r(e^{\lambda_n}E_n))),
\end{align*}
\begin{align*}
\lefteqn{\partial_{w}G_{n}^{\pm}(t, r, w, L) = (e^{\mu_{n} -
\lambda_{n}} \frac{1+L/t^{2}}{(1+w^{2}+L/t^{2})^{3/2}},} \\
& & {} & & {} & & {} & & {}- e^{\mu_{n} - \lambda_{n}}
\tilde{\mu}_{n} \frac{w}{\sqrt{1+w^{2}+L/t^{2}}} -
\dot{\lambda}_{n}),
\end{align*}
and thus
\begin{align*}
\mid \partial_{r}G_{n}^{\pm}(t, r, w, L) \mid & \leq C_{1}(t)
(c_{1}
+ D_{n}(t)+\Delta_{n-1}(t)+\Delta_{n}(t)),\\
\mid \partial_{w}G_{n}^{\pm}(t, r, w, L) \mid & \leq C_{1}(t),
\end{align*}
for $t \in [t_{0},T^{(1)}[$, $r \in \mathbb{R}$, $L \in [0,
L_{0}]$ and $\mid w \mid \leq z_{1}(t)$. Differentiating the
characteristic system with respect to $r$, we obtain
\begin{align*}
\frac{d}{ds}\partial_{r}(R_{n+1}^{\pm}, W_{n+1}^{\pm})(s, t, r, w,
L) =&
\partial_{r}G_{n}^{\pm}(s, R_{n+1}^{\pm}, W_{n+1}^{\pm}, 
L).\partial_{r}R_{n+1}^{\pm}(s, t, r, w, L)\\&+\partial_{w}G_{n}^{\pm}(s, R_{n+1}^{\pm}, W_{n+1}^{\pm}, 
L).\partial_{r}W_{n+1}^{\pm}(s, t, r, w, L),
\end{align*}
it follows that
\begin{align*}
\mid \frac{d}{ds}\partial_{r}(R_{n+1}^{\pm}, W_{n+1}^{\pm})(s, t,
r, w, L)\mid & \leq C_{1}(s) (c_{1} +
D_{n}(s)+\Delta_{n-1}(s)+\Delta_{n}(s))\\
& \ \mid
\partial_{r}(R_{n+1}^{\pm}, W_{n+1}^{\pm})(s, t, r, w, L)\mid,
\end{align*}
therefore by Gronwall's inequality we obtain,\\ for $(r, w, L) \in
{\rm supp} f^{+}_{n+1}(t) \cup {\rm supp} f^{+}_{n}(t) \cup {\rm supp} 
f^{-}_{n+1}(t) \cup {\rm supp} f^{-}_{n}(t)$
\begin{eqnarray*}
\mid \partial_{r}(R_{n+1}^{\pm}, W_{n+1}^{\pm})(t_{0}, t, r, w,
L)\mid \leq \exp\left[\int_{t_0}^{t}C_{1}(s) (c_{1} +
D_{n}(s)+\Delta_{n-1}(s)+\Delta_{n}(s)) ds\right]
\end{eqnarray*}
The definition of $f_{n}^{\pm}$ implies that
\begin{eqnarray*}
\parallel \partial_{r}f_{n}^{\pm}(t) \parallel \leq \parallel
\partial_{(r,w)} \overset{\circ}{f^{\pm}} \parallel \sup \{
\mid \partial_{r}(R_{n}^{\pm}, W_{n}^{\pm})(t_{0}, t, r, w, L)\mid
| (r, w, L) \in {\rm supp} f_{n}^{+}(t)\cup {\rm supp}
f_{n}^{-}(t) \}.
\end{eqnarray*}
Combining this with the previous inequality and using the
definition of $D_{n}$ we obtain the following :
\begin{eqnarray}\label{eq:3.19}
D_{n+1}(t) \leq (\parallel
\partial_{(r,w)} \overset{\circ}{f^{+}} \parallel+\parallel
\partial_{(r,w)} \overset{\circ}{f^{-}} \parallel)
\exp \left[\int_{t_0}^{t}C_{1}(s) (c_{1} +
D_{n}(s)+\Delta_{n-1}(s)+\Delta_n(s)) ds\right].
\end{eqnarray}
Let $\tilde{D}_{n}(t) := \sup \{ D_{m}(t) | m \leq n \}$ and
$\tilde{\Delta}_{n}(t) := \sup \{ \Delta_{m}(t) | m \leq n \}$.
Then $(\tilde{D}_{n})_{n}$ and $(\tilde{\Delta}_{n})_{n}$ are
increasing, therefore adding (\ref{eq:3.18}) and (\ref{eq:3.19})
implies
\begin{align*}
\tilde{D}_{n+1}(t)+\tilde{\Delta}_{n}(t) & \leq
C_1(t)+(1+\parallel
\partial_{(r,w)} \overset{\circ}{f^{+}} \parallel+\parallel
\partial_{(r,w)} \overset{\circ}{f^{-}} \parallel)\\
& \ \ \exp \left[\int_{t_0}^{t}C_{1}(s) (c_{1} +
\tilde{D}_{n+1}(s)+\tilde{\Delta}_n(s)) ds\right].
\end{align*}
Let $z_2$ be the right maximal
solution of
\begin{eqnarray*}
z_2(t) = C_1(t)+(1+\parallel
\partial_{(r,w)} \overset{\circ}{f^{+}} \parallel+\parallel
\partial_{(r,w)} \overset{\circ}{f^{-}} \parallel)
\exp \left[\int_{t_0}^{t}C_{1}(s) (c_{1} + z_2(s)) ds\right],
\end{eqnarray*}
which exists on an interval $[t_0, T^{(2)}[\subset[t_0, T^{(1)}[$. Then it follows that $$\tilde{D}_{n+1}(t)+\tilde{\Delta}_{n}(t) \leq z_2(t), \
t\in[t_0, T^{(2)}[, \ n\in \mathbb{N}.$$
Therefore there exists a
continuous function $C_2(t)$ which depends only on $z_2$ as an
increasing function such that all the quantities estimated against
$D_n$ and $\Delta_n$ are bounded in the $L^{\infty}$-norm by
$C_2(t)$. $\Box$

The following lemma deals with convergence of iterates. 
\begin{lemma}\label{l3}
The sequences $f_{n}^{+}$, $f_{n}^{-}$, $\lambda_n$, $\mu_n$, $E_n$, 
$\dot{\lambda}_n$, $\dot{\mu}_n$, $\tilde{\mu}_n$, $\rho_n$, $p_n$, 
$j_n$, $a_n$, $b_n$ converge uniformly on every compact subset 
$[t_0, T^{(3)}] \subset [t_0, T^{(2)}[$ on which the previous estimates hold.
\end{lemma}
{\bf Proof} Define for $t \in [t_0, T^{(3)}]$
\begin{eqnarray*}
\alpha_n(t):=\sup\left\{\parallel 
(f^{+}_{n+1}-f^{+}_{n})(s)\parallel+\parallel 
(f^{-}_{n+1}-f^{-}_{n})(s)\parallel+\parallel 
(\lambda_{n+1}-\lambda_{n})(s)\parallel \right.{}\\
\left. {}
 +\parallel (\mu_{n+1}-\mu_{n})(s)\parallel +\parallel 
(e^{\lambda_{n}}E_n-e^{\lambda_{n-1}}E_{n-1})(s)\parallel; \ t_0 \le 
s\le t\right\}
\end{eqnarray*}
and let $C$ denote a constant which may depend on the functions $z_1$ 
and $z_2$ introduced previously. Then using Lemma \ref{l1}, we have
\begin{eqnarray*}
\parallel \rho_{n+1}(t)-\rho_{n}(t)\parallel, \ \parallel 
\rho_{n+1}(t)-\rho_{n}(t)\parallel, \ \parallel 
p_{n+1}(t)-p_{n}(t)\parallel, \\
\parallel j_{n+1}(t)-j_{n}(t)\parallel, \ \parallel 
a_{n+1}(t)-a_{n}(t)\parallel, \ \parallel b_{n+1}(t)-b_{n}(t)\parallel 
\le C \alpha_n(t).
\end{eqnarray*}
Using mean value theorem to estimate differences 
$e^{\lambda_{n+1}+\mu_{n+1}}-e^{\lambda_{n}+\mu_{n}}$ and 
$e^{\mu_{n+1}}-e^{\mu_{n}}$, we 
deduce from (\ref{eq:3.7}) and (\ref{eq:3.9}) that
$$\parallel \dot{\lambda}_{n+1}(t)-\dot{\lambda}_{n}(t)\parallel, \ 
\parallel \tilde{\mu}_{n+1}(t)-\tilde{\mu}_{n}(t)\parallel \le C 
\alpha_n(t),$$ (\ref{eq:3.8}) thus implies
\begin{equation}\label{eq:3.20}
\parallel (\lambda_{n+1}-\lambda_{n})(t)\parallel \le C 
\int^{t}_{t_0}\alpha_n(s) ds.
\end{equation}
By mean value theorem (\ref{eq:3.6}) implies
\begin{equation}\label{eq:3.21}
\parallel (\mu_{n+1}-\mu_{n})(t)\parallel \le C 
\int^{t}_{t_0}\alpha_n(s) ds.
\end{equation}
Reasoning as in step 3 for the proof of theorem 3.1 in \cite{rein1} we can prove that
$$|(R,W)_{n+1}^{+}-(R,W)_{n}^{+}|(t_0, t, r, w, L), \ 
|(R,W)_{n+1}^{-}-(R,W)_{n}^{-}|(t_0, t, r, w, L) \le C 
\int_{t_0}^{t}\alpha_{n-1}(s) 
ds$$ which implies, using the fact that $f_{n}^{\pm}$ was defined in 
terms 
of the characteristics, and by mean value theorem
\begin{equation}\label{eq:3.22}
\parallel (f_{n+1}^{+}-f_{n}^{+})(t)\parallel, \ \parallel 
(f_{n+1}^{-}-f_{n}^{-})(t)\parallel \le C \int^{t}_{t_0}\alpha_{n-1}(s) 
ds.
\end{equation}
From (\ref{eq:3.10}) we deduce that
\begin{equation}\label{eq:3.23}
\parallel (e^{\lambda_n}E_{n}-e^{\lambda_{n-1}}E_{n-1})(t)\parallel \le 
C \int^{t}_{t_0}\alpha_{n-1}(s) ds.
\end{equation}
Adding (\ref{eq:3.20})-(\ref{eq:3.23}) gives
$$\alpha_n(t) \le C \int_{t_0}^{t}(\alpha_n(s)+\alpha_{n-1}(s)) ds,$$ 
then by Gromwall's inequality $$\alpha_n(t) \le C 
\int_{t_0}^{t}\alpha_{n-1}(s) ds,$$ and by induction $$\alpha_n(t) \le 
\frac{C^{n+1}}{n!}, \ 
\textrm{for} \ n \in \mathbb{N}, \ t \in [t_0, T^{(3)}].$$ We then 
conclude the uniform convergence of the iterates. In $L^\infty$-norm, 
$\lambda_n \to \lambda$, $\mu_n \to \mu$, $\tilde{\mu}_n \to 
\tilde{\mu}$, 
$f^{+}_n \to f^{+}$, $f^{-}_n \to f^{-}$, $E_n \to E$. $\Box$

In order to prove that the latter limits are regular in the sense of 
Definition \ref{d1}, we need to show the uniform convergence of the 
derivatives of the iterates above.
\begin{lemma}\label{l4}
The sequences $\partial_r f^{+}_n$, $\partial_w f^{+}_n$, $\partial_r 
f^{-}_n$, $\partial_w f^{-}_n$, $\lambda'_n$, $\mu'_n$, 
$\tilde{\mu}'_n$, $\partial_r (e^{\lambda_n}E_n)$ are uniformly convergent on $[t_0,T^{(3)}]$.
\end{lemma}
{\bf Proof} Fix $t \in [t_0, T^{(3)}]$, 
$|w|<U$, $L<\min(L_{0}^{+}, L_{0}^{-})$, $t_0 \le s\le t$. For 
$\partial 
\in \{ \partial_{r}, \partial_{w} \}$ and $s \mapsto (R_{n}^{\pm}(s), 
W_{n}^{\pm}(s))$ the solution of the characteristic system associated 
to 
equation (\ref{eq:3.5}) in $f_{n}^{\pm}$, define
\begin{align}\label{eq:3.24}
\xi_{n}^{\pm}(s):= e^{(\lambda_n - \mu_n)(s, R_{n}^{\pm})}\partial 
R_{n}^{\pm}(s, t, r, w,L),
\end{align}
\begin{align} \label{eq:3.25}
\eta_{n}^{\pm}(s) & := \partial W_{n}^{\pm} (s, t, r, w, L)  \notag\\
 & + (\sqrt{1+w^{2}+L/s^{2}} e^{\lambda_n - \mu_n}
\dot{\lambda}_n)|_{(s, (R_{n}^{\pm}, W_{n}^{\pm})(s, t, r, w, L ))}
\partial R_{n}^{\pm}(s, t, r, w,L).
\end{align} 
 Reasoning as in step 4 for the proof of theorem 3.1 in \cite{rein1} it 
can be proved with minor changes that for all 
$\epsilon>0$, there is a non negative integer $N$ such that for $n>N$, 
\begin{align}\label{eq:3.26}
(\mid \xi_{n+1}^{\pm} - \xi_{n}^{\pm}\mid + \mid \eta_{n+1}^{\pm} -
 \eta_{n}^{\pm}\mid)(s)& \leq C \epsilon \notag \\
&+ C \int_{t_0}^{s}(\mid \xi_{n+1}^{\pm} -
\xi_{n}^{\pm}\mid + \mid \eta_{n+1}^{\pm} - \eta_{n}^{\pm}\mid)(\tau) 
d\tau.
\end{align}
By Gronwall's lemma it follows that the sequences $\xi_{n}^{\pm}$ and 
$\eta_{n}^{\pm}$ converge uniformly. The transformation from $(\partial 
R_{n}^{\pm}, \partial W_{n}^{\pm})$ to $(\xi_{n}^{\pm}, 
\eta_{n}^{\pm})$ being invertible with convergent coefficients, we 
deduce the 
convergence of $\partial_{r,w} (R_{n}^{\pm}, W_{n}^{\pm})$ and thus the 
convergence of $\partial_r f_{n}^{\pm}$ and $\partial_w f_{n}^{\pm}$, 
using 
the fact that $f_{n}^{\pm}$ was defined in terms of characteristics.\\ 
Let us prove the convergence of $\lambda'_{n}$, $\mu'_{n}$, 
$\tilde{\mu}'_{n}$, $\partial_r(e^{\lambda_{n}}E_n)$. Define
\begin{align}\label{eq:3.27}
\gamma_n(t)& :=\sup\{\mid \xi_{n+1}^{+} -
\xi_{n}^{+}\mid(s)+\mid \xi_{n+1}^{-} -
\xi_{n}^{-}\mid(s) + \mid \eta_{n+1}^{+} - \eta_{n}^{+}\mid(s)\notag \\
& +\mid \eta_{n+1}^{-} - \eta_{n}^{-}\mid(s) 
+\parallel(\mu'_{n+1}-\mu'_{n})(s)\parallel+\parallel(\lambda'_{n+1}-\lambda'_{n})(s)\parallel 
\notag \\
& +\parallel 
\partial_r(e^{\lambda_{n+1}}E_{n+1})-\partial_r(e^{\lambda_{n}}E_{n})\parallel(s), 
t_0 \le s \le t \}.
\end{align}
The sequences $\mu_n$, $\dot{\mu}_{n}$, $\tilde{\mu}_n$, 
$\dot{\lambda}_n$, $e^{\lambda_n}E_n$, $\rho_n$, $j_n$, $a_n$, $b_n$ 
converge 
uniformly, we then take the above integer $N$ large enough so that we 
have for 
$n>N$,
\begin{align} \label{eq:3.28}
\parallel & 
\left(e^{\lambda_{n}}E_{n}-e^{\lambda_{n-1}}E_{n-1}\right)(s)\parallel, 
\parallel(\mu_{n+1}-\mu_{n})(s)\parallel, 
\parallel(\tilde{\mu}_{n}-\tilde{\mu}_{n-1})(s)\parallel,  \notag \\
&\parallel(\dot{\lambda}_{n+1}-\dot{\lambda}_{n})(s)\parallel, 
\parallel(\rho_{n+1}-\rho_{n})(s)\parallel, 
\parallel(j_{n+1}-j_{n})(s)\parallel, \notag \\
&\parallel(\dot{\mu}_{n+1}-\dot{\mu}_{n})(s)\parallel, 
\parallel(a_{n+1}-a_{n})(s)\parallel, 
\parallel(b_{n+1}-b_{n})(s)\parallel\le \epsilon.
\end{align}
Taking $\partial=\partial_r$, it follows from 
(\ref{eq:3.24})-(\ref{eq:3.25}) that
\begin{align}\label{eq:3.29}
\partial R_{n}^{\pm}(s):= e^{(\mu_n - \lambda_n)(s, 
R_{n}^{\pm})}\xi_{n}^{\pm}(s),
\end{align}
\begin{align} \label{eq:3.30}
\partial W_{n}^{\pm} (s) = \eta_{n}^{\pm}(s)  
  - (\sqrt{1+w^{2}+L/s^{2}} 
\dot{\lambda}_n)\xi_{n}^{\pm}(s).
\end{align}
Using these equations and since $f_{n}^{\pm}$ was defined in terms of 
characteristics we obtain
\begin{align}\label{eq:3.31}
\parallel (\partial_rf_{n+1}^{\pm}-\partial_rf_{n}^{\pm})(s)\parallel & 
\leq \parallel \partial_{r,w}\overset{\circ}{f}^{\pm}\parallel(| 
\partial_rR_{n+1}^{\pm}-\partial_rR_{n}^{\pm}|+|\partial_rW_{n+1}^{\pm}-\partial_rW_{n}^{\pm}|)(s) 
\notag \\
& \le \parallel 
\partial_{r,w}\overset{\circ}{f}^{\pm}\parallel(|e^{\mu_{n+1}-\lambda_{n+1}}\xi_{n+1}^{\pm}-e^{\mu_{n}-\lambda_{n}}\xi_{n}^{\pm}| 
\notag \\
& 
+|\eta_{n+1}^{\pm}-\eta_{n}^{\pm}|+|\dot{\lambda}_{n+1}\xi_{n+1}^{\pm}-\dot{\lambda}_{n}\xi_{n}^{\pm}|)(s)
\end{align}
this implies, using (\ref{eq:3.28}) and the fact that $\lambda_n$, 
$\mu_n$, $\dot{\lambda}_n$, $\xi_n$ are bounded:
\begin{align}\label{eq:3.32}
\parallel (\partial_rf_{n+1}^{\pm}-\partial_rf_{n}^{\pm})(s)\parallel  
\leq 
C\epsilon+C(|\xi_{n+1}^{\pm}-\xi_{n}^{\pm}|+|\eta_{n+1}^{\pm}-\eta_{n}^{\pm}|)(s).
\end{align}
Using (\ref{eq:3.27}), (\ref{eq:3.28}), (\ref{eq:3.32}) and the 
expressions of $p_n$, $\rho_n$, $j_n$, we deduce that
\begin{equation}\label{eq:3.33}
\parallel(\rho'_{n+1}-\rho'_{n})(s)\parallel, 
\parallel(p'_{n+1}-p'_{n})(s)\parallel, 
\parallel(j'_{n+1}-j'_{n})(s)\parallel \le 
C\epsilon+C(\gamma_n+\gamma_{n-1})(s). 
\end{equation}
Now taking the derivative of (\ref{eq:3.6}) with respect to $r$, using 
(\ref{eq:3.28}), (\ref{eq:3.33}) and the fact that $\mu_n$, $\mu'_{n}$ 
are bounded we deduce an estimate for $\mu'_{n}$:
\begin{equation}\label{eq:3.34}
\parallel(\mu'_{n+1}-\mu'_{n})(s)\parallel \le 
C\epsilon+C\int_{t_0}^{s}(\gamma_n+\gamma_{n-1})(\tau) d\tau. 
\end{equation}
For $\lambda'_{n}$, we first take the derivative of (\ref{eq:3.7}) with 
respect to $r$ and obtain
\begin{equation}\label{eq:3.35}
\dot{\lambda}'_{n}=(8\pi t \mu'_{n}\rho_n+4\pi t 
\rho'_{n})e^{2\mu_n}+(\Lambda t-k/t)\mu'_{n}e^{2\mu_n}, 
\end{equation}
this shows that $\dot{\lambda}'_{n}$ is bounded. Subtracting 
(\ref{eq:3.35}) written for $n+1$ and $n$, we obtain the following, 
using the 
fact that $\rho_n$, $\mu_n$, $\mu'_{n}$ are bounded and 
(\ref{eq:3.27}), 
(\ref{eq:3.28}) and (\ref{eq:3.32}):
\begin{equation}\label{eq:3.36}
\parallel(\dot{\lambda}'_{n+1}-\dot{\lambda}'_{n})(s)\parallel \le 
C\epsilon+C(\gamma_n+\gamma_{n-1})(s), 
\end{equation}
and integrating this over $[t_0, t]$ it follows from (\ref{eq:3.8}) 
that
\begin{equation}\label{eq:3.37}
\parallel(\lambda'_{n+1}-\lambda'_{n})(s)\parallel \le 
C\epsilon+C\int_{t_0}^{t}(\gamma_n+\gamma_{n-1})(\tau)d\tau. 
\end{equation}
For $\tilde{\mu}'_{n}$, we take the derivative of (\ref{eq:3.9}) with 
respect to $r$, subtract the expressions written for $n+1$ and $n$, and 
use (\ref{eq:3.27}), (\ref{eq:3.28}) and (\ref{eq:3.32}) to obtain
\begin{equation}\label{eq:3.38}
\parallel(\tilde{\mu}'_{n+1}-\tilde{\mu}'_{n})(s)\parallel \le 
C\epsilon+C(\gamma_n+\gamma_{n-1})(s). 
\end{equation}
Now for an estimate for $\partial_r(e^{\lambda_n}E_n)$,
we subtract the expressions written for $n+1$ and $n$ from 
(\ref{eq:3.11}), use (\ref{eq:3.27}), (\ref{eq:3.28}), (\ref{eq:3.32}) 
and the fact 
that $\lambda_n$, $\mu_n$, $\mu'_{n}$, $a_n$, $b_n$,  $\tilde{\mu}_n$ 
are bounded to obtain
\begin{equation}\label{eq:3.39}
\parallel(\partial_r(e^{\lambda_{n+1}}E_{n+1})-\partial_r(e^{\lambda_n}E_n))(s)\parallel 
\le C\epsilon+C\int_{t_0}^{t}(\gamma_n+\gamma_{n-1})(\tau)d\tau. 
\end{equation}
Now we can add (\ref{eq:3.26}), (\ref{eq:3.34}), (\ref{eq:3.37}) and 
(\ref{eq:3.39}) to obtain, after taking the supremum over 
$s\in[t_0,t]$:
\begin{equation}\label{eq:3.40}
\gamma_n(t) \le C\epsilon+C\int_{t_0}^{t}(\gamma_n+\gamma_{n-1})(s)ds, 
\end{equation}
and setting $\tilde{\gamma}_n(t)=\sup\{\gamma_m, m \le n\}$, we deduce 
by Gronwall's lemma that
\begin{equation*}
\tilde{\gamma}_n(t) \le C\epsilon, \ n>N, \ t\in [t_0, T^{(3)}].
\end{equation*}
Thus the sequence $\tilde{\gamma}_n$ converges uniformly to $0$. By 
(\ref{eq:3.34}), (\ref{eq:3.37}), (\ref{eq:3.38}) and (\ref{eq:3.39}), 
the 
sequences $\tilde{\mu}'_{n}$, $\mu'_{n}$, $\lambda'_{n}$ and 
$\partial_r(e^{\lambda_n}E_n)$ then converge uniformly on $[t_0, 
T^{(3)}]$. 
$\Box$

The regularity of $f^{+}$, $f^{-}$, $\lambda$, $\mu$, $\tilde{\mu}$ and 
$E$ follows. Using the convergence of the derivatives, it can be proven 
as in \cite{rein1} that $(f^+, f^-, \lambda, \mu, \tilde{\mu}, E)$ is a 
regular solution of (\ref{eq:3.1}), (\ref{eq:3.2}), (\ref{eq:1.4}), 
(\ref{eq:1.5}), (\ref{eq:3.3}), (\ref{eq:1.9}) and then
$(f^+,f^-,\lambda,\mu,E)$ solves (\ref{eq:1.2})-(\ref{eq:1.9}). To end 
the proof of Theorem \ref{local}, we prove the uniqueness of the 
solution. Let $(f^{+}_{i},f^{-}_{i},\lambda_{i},\mu_{i},E_{i})$, $i=1, 
2$ be 
two regular solutions of the Cauchy problem for the same initial data 
$(\overset{\circ}{f^+},\overset{\circ}{f^-},
\overset{\circ}{\lambda},
\overset{\circ}{\mu},\overset{\circ}{E})$ at $t=t_0$. Setting 
\begin{eqnarray*}
\alpha(t):=\sup\left\{\parallel 
(f^{+}_{1}-f^{+}_{2})(s)\parallel+\parallel 
(f^{-}_{1}-f^{-}_{2})(s)\parallel+\parallel 
(\lambda_{1}-\lambda_{2})(s)\parallel \right.{}\\
\left. {}
 +\parallel (\mu_{1}-\mu_{2})(s)\parallel +\parallel 
(e^{\lambda_{1}}E_1-e^{\lambda_{2}}E_{2})(s)\parallel; \ t_0 \le s\le 
t\right\},
\end{eqnarray*}
and proceeding similarly as to prove the convergence of iterates leads 
to $$\alpha(t)\le C \int_{t_0}^{t}\alpha(s) ds,$$ which implies that 
$\alpha(t)=0$ for $t\in[t_0, \infty[$. This proves uniqueness and 
completes the proof of Theorem \ref{local}.
\subsection{Proof of Theorem \ref{cont}}
Let $(f^+, f^-, \lambda, \mu, E)$ be a right maximal 
solution of the full system (\ref{eq:1.2})-(\ref{eq:1.9}) with existence interval $[t_0, T_{max}[$. We 
assume that $T_{max}<\infty$. By assumption 
\begin{align*} 
Q_{*} &:= \sup \{ te^{2\mu(t,r)}|r \in \mathbb{R}, t_0\leq
t< T_{max} \}<\infty,\\
S_{*} &:= \sup \{ |E|e^{\lambda(t,r)}|r \in
\mathbb{R}, t_0\leq t< T_{max} \}<\infty,
\end{align*}
and $P_{*}<\infty$ where $P_{*}:=P_{*}^{+}+P_{*}^{-}$ with 
$$P_{*}^{\pm}:=\sup\{|w|, (r,w,L)\in\supp f^{\pm}(t), t\in[t_0, 
T_{max}[\}.$$  We take $t_1\in]t_0, T_{max}[$, and we will 
show that the system has a solution with initial data $(f^+(t_1), 
f^-(t_1), \lambda(t_1), \mu(t_1), E(t_1))$ prescribed at $t=t_1$ which 
exists 
on an interval $[t_1, t_1+\delta]$ with $\delta >0$ independent of 
$t_1$. By moving $t_1$ close enough to $T_{max}$ this would extend the 
initial solution beyond $T_{max}$, a contradiction to the initial 
solution 
being right maximal. We have proved previously that such a solution 
exists at least on the right maximal existence interval of the 
solutions 
$z_1$, $z_2$ of 
\begin{equation*}
    z_1(t) = W(t_1)+\|t_{1}e^{2\mu(t_1)}\|+
    \|e^{\lambda(t_1)}E(t_1)\|+C^*\int_{t_1}^{t}
 (1+z_1(s))^{15/2} ds,
\end{equation*}
\begin{eqnarray*}
z_2(t) = C_1(t)+(1+\parallel
\partial_{(r,w)} f^{+}(t_1) \parallel+\parallel
\partial_{(r,w)} f^{-}(t_1) \parallel)
\exp \left[\int_{t_1}^{t}C_{1}(s) (c_{1} + z_2(s)) ds\right],
\end{eqnarray*}
where $W(t_1):=W^{+}(t_1)+W^{-}(t_1)$,
\begin{align*}
W^{\pm}(t_1)&:=\sup\{|w|, (r,w,L)\in\supp f^{\pm}(t_1)\},\\
C^*&=C(1+\Lambda)(1+L_0)^2(1+\|f^{+}(t_1)\|+\|f^{-}(t_1)\|),\\ 
c_1&:=1+\Lambda+\|\lambda'(t_1)\|+\|e^{-2\mu(t_1)}\mu'(t_1)\|,
\end{align*}
and $C_1$ is an increasing function of $z_1$. Now $W(t_1) \le P_*$, 
$\|t_{1}e^{2\mu(t_1)}\| \le Q_*$, $\|e^{\lambda(t_1)}E(t_1)\| \le S_*$, 
$\|f^{\pm}(t_1)\|=\|\overset{\circ}{f^{\pm}}\|$, $L_0$ is unchanged. 
Thus 
we have uniform bounds 
$W(t_1)+\|t_{1}e^{2\mu(t_1)}\|+\|e^{\lambda(t_1)}E(t_1)\|\le M_1$, 
$C^*\le M_2$. On the other hand we can use the 
expressions for $\lambda'$, $\mu'$, $\dot{\lambda}'$, some estimates 
proved 
in lemmas 3.1, 3.2, and 3.3 to obtain uniform bounds $c_1\le M_3$, 
$C_1(t)+(1+\parallel
\partial_{(r,w)} f^{+}(t_1) \parallel+\parallel
\partial_{(r,w)} f^{-}(t_1) \parallel) \le M_4$. Let $y_1$ and $y_2$ be 
the right maximal solution of
\begin{align*}
y_1(t) &= M_1+M_2\int_{t_1}^{t}
 (1+y_1(s))^{15/2} ds,\\
y_2(t) &= M_4
\exp \left[\int_{t_1}^{t}C_{1}^{*}(s) (M_3 + y_2(s)) ds\right],
\end{align*}
respectively, where $C_{1}^{*}$ depends on $y_1$ in the same way as 
$C_1$ depends on $z_1$. Then $y_1$ and $y_2$ exist on an interval 
$[t_1, 
t_1+\delta]$ with $\delta >0$ independent of $t_1$. If we choose $t_1$ 
such that $T_{max}< t_1+\delta$ then $z_1\le y_1$, $z_2\le y_2$, in 
particular $z_1$ and $z_2$ exist on $[t_1, t_1+\delta]$. This completes 
the 
proof of theorem \ref{cont}.

\subsection{Proof of Proposition 1.3}
We start by showing how to obtain the bound on $w$. Since we are in the 
non-vacuum case one has $w_{0}^{\pm}>0$ and $L_{0}^{\pm} >0$. For $t 
\geq 
t_0$ define
\begin{align*}
& P_{+}^{\pm}(t) := \max \{0, \max \{w | (r,w,L) \in {\rm supp} 
f^{\pm}(t) \} \},\\
& P_{-}^{\pm}(t) := \min \{0, \min \{ w | (r,w,L) \in {\rm supp} 
f^{\pm}(t) \}
 \}.
\end{align*}
Let $(r(s),w(s), L)$ be a characteristic curve in the support of 
$f^{\pm}$. Assume that $P_{+}^{\pm}(t)>0$ for some $t\in[t_0, 
T_{max}[$, and 
let $w(t)>0$. Let $t_1 \in [t_0,t[$ be defined minimal such 
that 
$w(s)>0$ for $s\in [t_1,t[$. We have
\begin{align}\label{eq:3.41}
\dot{w}(s) &= -\dot{\lambda}w-e^{\mu-\lambda}\mu'\sqrt{1+w^2+L/s^2} \mp 
e^{\lambda+\mu}E \nonumber\\
& = \frac{4
\pi^{2}}{s}e^{2\mu}\int_{-\infty}^{\infty}\int_{0}^{\infty}
\left(\tilde{w}\sqrt{1+w^{2}+L/s^{2}}-w\sqrt{1+\tilde{w}^{2}+\tilde{L}/s^{2}}\right)(f^{+}+f^-)
d\tilde{L}d\tilde{w} \nonumber\\
& \mp e^{\lambda+\mu}E+\frac{1}{2s}w+\frac{(k-\Lambda 
s^2)e^{2\mu}}{2s}w
-2\pi s e^{2\mu}(e^{2\lambda}E^2+cs^{-4})w.
\end{align}
As long as $w(s)>0$ we drop the last two terms of 
the right hand side in (\ref{eq:3.41}) since they are negative, and 
then obtain
\begin{align}\label{eq:3.42}
\dot{w}(s) &\le \frac{C}{s}\int_{0}^{P_{+}^{\pm}(s)}\int_{0}^{L_0}
\frac{\tilde{w}(1+L)}{w}(f^{+}+f^-)
d\tilde{L}d\tilde{w} + \frac{w}{s}+ 
e^{\lambda+\mu}|E|\nonumber\\
& \le \frac{C}{s}\left[\frac{(P_{+}^{\pm}(s))^{2}}{w(s)}+w(s)\right]+ 
e^{\lambda+\mu}|E|.
\end{align}
Now integrating (\ref{eq:1.9}) with respect to $t$, using the boundedness of $\mu$ and the fact that $s^2|b(s)|\leq s^2 \bar{a}(s) \leq C P_{+}^{\pm}(s)$ ($\bar{a}$ being similar to $a$ with $f^+-f^-$ replaced by $f^++f^-$), we obtain
$$(e^{\lambda+\mu}|E|)(s)\leq Ct_{1}^{2}(e^{\lambda}|E|)(t_1)+C\int_{t_1}^{s}P_{+}^{\pm}(\tau)d\tau.$$
It then follows that
$$\frac{d}{ds}w(s)^{2}\le \frac{C}{s}(P_{+}^{\pm}(s))^{2}+C 
P_{+}^{\pm}(s)+C P_{+}^{\pm}(s)\int_{t_1}^{s}P_{+}^{\pm}(\tau)d\tau,$$ which implies after integration 
$$w^{2}(t)\le 
w^{2}(t_1)+C\int_{t_1}^{t}\left[s^{-1}(P_{+}^{\pm}(s))^{2}+P_{+}^{\pm}(s)\right]ds+ C\int_{t_1}^{t}\int_{t_1}^{s}P_{+}^{\pm}(s)P_{+}^{\pm}(\tau)d\tau ds.$$ 
If $t_1=t_0$ then $w(t_1)\le w_0$, 
otherwise $w(t_1)=0$. In any case it follows that 
\begin{equation}\label{w}
w^{2}(t)\le 
w^{2}_{0}+C\int_{t_1}^{t}\left[s^{-1}(P_{+}^{\pm}(s))^{2}+P_{+}^{\pm}(s)\right]ds+ C\int_{t_1}^{t}\int_{t_1}^{s}P_{+}^{\pm}(s)P_{+}^{\pm}(\tau)d\tau ds.
\end{equation} 
The double integral in the right hand side of (\ref{w}) needs to be worked out.
\begin{align*}
\int_{t_1}^{t}\int_{t_1}^{s}P_{+}^{\pm}(s)P_{+}^{\pm}(\tau)d\tau ds&\leq\frac{1}{2}\int_{t_1}^{t}\int_{t_1}^{s}(P_{+}^{\pm}(s))^2d\tau ds+\frac{1}{2}\int_{t_1}^{t}\int_{t_1}^{s}(P_{+}^{\pm}(\tau))^2d\tau ds\\&=\frac{1}{2}\int_{t_1}^{t}(s-t_1)(P_{+}^{\pm}(s))^2 ds+\frac{1}{2}\int_{t_1}^{t}\int_{\tau}^{t}(P_{+}^{\pm}(\tau))^2ds d\tau\\&\leq Ct\int_{t_1}^{t}(P_{+}^{\pm}(s))^2ds.  
\end{align*}
Therefore (\ref{w}) implies
$$(P_{+}^{\pm}(t))^{2}\le 
(w^{2}_{0}+Ct)+C\int_{t_1}^{t}(s^{-1}+t+1)(P_{+}^{\pm}(s))^{2} ds, \ 
\rm{for} \ t\in[t_0, T_{max}[.$$ By 
Gronwall's inequality it follows that $P_{+}^{\pm}$ is bounded on $[t_0, T_{max}[$.\\ Estimating $\dot{w}(s)$ from below in the case $w(s)<0$ along the same lines shows that $P_{-}^{\pm}$ is bounded as well.

The bounds on $w$ and $\mu$ imply that $e^\lambda |E|$ is bounded as well, using (\ref{eq:1.9}).

\subsection{Proof of Theorem \ref{global}}
We prove that $\mu$ is bounded on $[t_0, T_{max}[$.\\A lengthy computation leads to
\begin{align}\label{eq:3.43}
\frac{d}{dt}\int_{0}^{1} e^{\mu+\lambda}\rho(t,r) dr &=
-\frac{1}{t}\int_{0}^{1} e^{\mu+\lambda}\left[2\rho + q
-\frac{\rho+p}{2}[1+(k-\Lambda t^{2}) e^{2\mu}]\right] dr \nonumber\\
&\leq-\frac{2}{t}\int_{0}^{1} e^{\mu+\lambda}\rho 
dr+\frac{1}{t}\int_{0}^{1} 
e^{\mu+\lambda}\frac{\rho+p}{2}[1+(k-\Lambda t^{2}) e^{2\mu}]dr,
\end{align}
since $q$ is nonnegative. Using the fact that $\rho+p \geq 0$, $\rho \geq p$  and $k-\Lambda t^2 \leq 0$ it follows that
\begin{align*}
\frac{d}{dt}\int_{0}^{1} e^{\mu+\lambda}\rho(t,r) dr\leq-\frac{1}{t}\int_{0}^{1} e^{\mu+\lambda}\rho 
dr,
\end{align*}
and by Gronwall's inequality 
\begin{align}\label{eq:3.44}
\int_{0}^{1} e^{\mu+\lambda}\rho(t,r) dr \leq Ct^{-1} \ t\in[t_0, T_{max}[.
\end{align}
On the other hand using the equation $\mu'=-4\pi t e^{\mu+\lambda} j$, (\ref{eq:3.44}) and the fact that $|j|\leq \rho$ we find
\begin{align}\label{eq:3.45}
|\mu(t,r)-\int_{0}^{1} \mu(t,\sigma) d\sigma|\leq C, \ t\in[t_0, T_{max}[, \ r\in[0,1].
\end{align}
Next using (\ref{eq:1.5}), $p-\rho \leq0$ and $k-\Lambda t^2 \leq 0$, we have
\begin{align*}
 \frac{ \partial}{\partial t}e^{\mu-\lambda} & =
 e^{\mu-\lambda}\left[4\pi
te^{2\mu}(p-\rho)+\frac{1+ke^{2\mu}}{t}-\Lambda t
 e^{2\mu}\right]\\
& \leq e^{\mu-\lambda}\left[\frac{1}{t}+\frac{k-\Lambda
t^2}{t}
 e^{2\mu}\right]\\
 & \leq  \frac{1}{t}e^{\mu-\lambda},
\end{align*}
so that
\begin{align}\label{eq:3.46}
e^{(\mu-\lambda)(t,r)}\leq C t, \ t\in[t_0, T_{max}[, \ r\in[0,1].
\end{align}
Now using $p-\rho \leq0$ and $k-\Lambda t^2 \leq 0$, it follows that
\begin{align*}
\int_{0}^{1}\mu(t,r) dr & = \int_{0}^{1}\overset{\circ}{\mu}(r) dr
+ \int_{t_0}^{t}\int_{0}^{1}\dot{\mu}(s,r) dr ds\\
& \leq C+\int_{t_0}^{t}\frac{1}{2s}\int_{0}^{1}[e^{2\mu}(8\pi
s^{2}p + k-\Lambda s^{2})+1]dr ds\\
& \leq C+\int_{t_0}^{t}\frac{1}{2s}\int_{0}^{1}(8\pi
s^{2}e^{\mu-\lambda}e^{\mu+\lambda}\rho + 1)dr ds\\
& \leq C+\frac{1}{2}\ln(t/t_{0})+C\int_{t_0}^{t}s^{2}\int_{0}^{1}e^{\mu+\lambda}\rho dr ds\\
& \leq C + \frac{1}{2}\ln(t/t_{0})+Ct^2,
\end{align*}
and using (\ref{eq:3.45}) we obtain
 \begin{align*}
\mu(t,r)\leq C(1+t^2+\ln t),\  t \in [t_0,T_{max}[ \ r\in[0,1].
\end{align*}
$\mu$ is then bounded on $[t_0,T_{max}[$ and by proposition 1.3 the 
proof of theorem \ref{global} is complete.
\subsection{Proof of Theorem \ref{com}}
The equation of motion for charged particles is given by the following differential system for a path $\nu \mapsto (\tau,v^0, v^i)(\nu)$:
\begin{align*}
\frac{d\tau}{d\nu}=v^0, \ \frac{dv^0}{d\nu}=k_{ij}v^iv^j, \ \frac{dv^i}{d\nu}=2k_{j}^{i}v^jv^0-\gamma^{i}_{mn}v^mv^n\mp(F_{0}\ ^{i}v^0+F_{j}\ ^{i}v^j).
\end{align*}
For a particle with rest mass $m$ moving forward in time, $v^0=(m^2+g_{ij}v^iv^j)^{1/2}$. Then the relation between coordinate time $\tau$ and proper time $\nu$ is 
\begin{equation}\label{gc}
\frac{d\tau}{d\nu}=(m^2+g_{ij}v^iv^j)^{1/2}. 
\end{equation}
In order to prove the completeness of trajectories it is useful to control $g_{ij}v^iv^j$ as a function of $\tau$. As in \cite{lee1}, we can define, from the Vlasov equations (\ref{vl1})-(\ref{vl2}), the characteristic curve $V^i(\tau)$ satisfying
\[
\frac{dV_i}{d\tau}=-(1+g_{rs}V^rV^s)^{-1/2}\gamma^{j}_{mn}V_pV_qg^{pm}g^{qn}g_{ij}\mp g_{ij}[F_{0}\ ^{j}+F_{l}\ ^{j}g^{lm}V_m(1+g_{rs}V^rV^s)^{-1/2}].
\]
In this section we use the notation $\gamma:=(\Lambda/3)^{1/2}$ and $C$ is an arbitrary positive constant which may change from line to line.\\ 
Using the latter equation we obtain
\begin{align}\label{eq:3.47}
\frac{d}{d\tau}(g^{ij}V_iV_j)&=2k^{ij}V_iV_j\mp 2g^{ij}V_iF_{0j} \nonumber\\ & \leq (-2\gamma+Ce^{-\gamma \tau})g^{ij}V_iV_j\mp 2g^{ij}V_iF_{0j}.
\end{align}
We have used \cite[(3.22)]{lee1}. The second term in the right hand side of (\ref{eq:3.47}) can be estimated using the Cauchy-Schwarz inequality and the elementary inequality $xy \leq \varepsilon x^2/2+y^2/2\varepsilon$ :
\begin{align}\label{eq:3.48}
g^{ij}V_iF_{0j}&\leq (g^{ij}V_iV_j)^{1/2}(g^{ij}F_{0i}F_{0j})^{1/2} \nonumber\\ & \leq \frac{\varepsilon}{2}(g^{ij}V_iV_j)+\frac{1}{2\varepsilon}(g^{ij}F_{0i}F_{0j}),
\end{align}
where $\varepsilon$ is such that $0<\varepsilon<2 \gamma$. On the other hand from the definition of $\tau_{00}$ it follows that 
\begin{equation}\label{eq:3.49}
g^{ij}F_{0i}F_{0j}\leq \tau_{00}.
\end{equation}
and using (\ref{eq:1.37}) $g^{ij}F_{0i}F_{0j}\leq C e^{-2\gamma \tau}$. Thus we deduce from (\ref{eq:3.47}) and (\ref{eq:3.48}) the following
\begin{align*}
\frac{d}{d\tau}(g^{ij}V_iV_j)\leq (-2\gamma+\varepsilon+Ce^{-\gamma \tau})g^{ij}V_iV_j+C e^{-\gamma \tau}.
\end{align*}
Setting $V:=e^{(2\gamma-\varepsilon)\tau}g^{ij}V_iV_j$, it follows that
\[
\frac{dV}{d\tau}\leq Ce^{-\gamma \tau}V+C e^{-\varepsilon \tau},
\]
$V$ is thus bounded by Gronwall inequality and then
\begin{align}\label{bd}
g^{ij}V_iV_j\leq C e^{(-2\gamma+\varepsilon)\tau}.
\end{align}
Therefore $g^{ij}V_iV_j$ is bounded.
This is enough to deduce from (\ref{gc}) that for $m\geq0$, we have 
\[
\frac{d\nu}{d\tau}\geq C,
\]
so that $\nu$ goes to infinity as does $\tau$. The completeness of causal trajectories is then proved.
\subsection{Proof of Theorem \ref{decay}} The proof is based on a bootstrap argument.

By hypothesis
\begin{equation*}
|t_0\dot{\lambda}(t_0)-1|\leq \delta,\ |(e^{-\lambda}\mu')(t_0)|\leq \delta,\ |(e^{\lambda}E)(t_0)|\leq \delta
\end{equation*}
\begin{equation*}
|\Lambda t_{0}^2e^{2\mu(t_0)}-3-3ke^{2\mu(t_0)}|\leq \delta,\ \bar{w}(t_0)\leq \delta,
\end{equation*}
and by continuity, this implies that
\begin{equation*}
|t\dot{\lambda}(t)-1|\leq 2\delta,\ |(e^{-\lambda}\mu')(t)|\leq 2\delta,\ |(e^{\lambda}E)(t)|\leq 2\delta
\end{equation*}
\begin{equation*}
|\Lambda t^2e^{2\mu(t)}-3-3ke^{2\mu(t)}|\leq 2\delta,\ \bar{w}(t)\leq 2\delta,
\end{equation*}
for $t$ close to $t_0$.

Let $C_1$ and $\epsilon$ be constants for $0<C_1<1$ and $0<\epsilon<1/2$. We can reduce $\delta$ if necessary so that $2\delta<C_1 t_{0}^{-3+\epsilon}$. Then there exists some time interval on which the following bootstrap assumption is satisfied
\begin{equation}\label{boot1}
|t\dot{\lambda}-1|\leq C_1t^{-2+\epsilon},\ |e^{-\lambda}\mu'|\leq C_1t^{-2+\epsilon},\ |e^{\lambda}E|\leq C_1t^{-2+\epsilon}
\end{equation}
\begin{equation}\label{boot2}
|\Lambda t^2e^{2\mu}-3-3ke^{2\mu}|\leq C_1t^{-3+\epsilon},\ \bar{w}(t)\leq C_1t^{-1+\epsilon}.
\end{equation}
Consider the maximal interval $[t_0,t_*)$ on which (\ref{boot1})-(\ref{boot2}) hold and suppose $t_*$ is finite.

Let us continue with the following set of equations:
\begin{align}
e^{-2\mu}(2t\dot{\lambda}+1)+k-\Lambda t^2=8\pi t^2 \rho \label{eq:3.52}\\
e^{-\lambda}\mu'=-4\pi t e^{\mu}j \label{eq:3.53}\\
\partial_t(t^2 e^\lambda
E)=-t^2 e^\mu b \label{eq:3.54}\\
e^{-2\mu}(2t\dot{\mu}-1)-k+\Lambda t^2=8\pi t^2 p\label{eq:3.55}\\
\dot{w}=-\dot{\lambda}w-e^{\mu-\lambda}\mu'\sqrt{1+w^2+L/t^2}\mp e^{\lambda+\mu}E. \label{eq:3.56}
\end{align}
From (\ref{eq:3.52}) we have
\begin{equation}\label{eq:3.57}
t\dot{\lambda}-1=\frac{1}{2}(\Lambda e^{2\mu}t^2-3-3k e^{2\mu})+k e^{2\mu}+4\pi t^2 e^{2\mu}\rho,
\end{equation}
from (\ref{eq:3.55})
\begin{equation}\label{eq:3.58}
\partial_t[-\frac{1}{3}t e^{-2\mu}(\Lambda t^2e^{2\mu}-3-3k e^{2\mu})]=-8\pi t^2 p,
\end{equation}
and from (\ref{eq:3.56})
\begin{equation}\label{eq:3.59}
\partial_t(tw)=-t^{-1}(t\dot{\lambda}-1)(tw)-t e^{\mu}(e^{-\lambda}\mu')\sqrt{1+w^2+L/t^2}\mp te^{\mu}(e^{\lambda}E).
\end{equation}
Consider a solution of the full system (\ref{eq:1.2})-(\ref{eq:1.15}) on the interval $[t_0,t_*)$  which satisfies the bootstrap assumption (\ref{boot1})-(\ref{boot2}). Putting inequalities (\ref{boot1})-(\ref{boot2}) into equations (\ref{eq:3.53})-(\ref{eq:3.54}), (\ref{eq:3.57})-(\ref{eq:3.59}) allows new estimates to be derived. For this purpose it is important to have estimates for the matter quantities $j$, $p$ and $b$.\\ Let $F:=\max\{\parallel f^+ \parallel_{L^{\infty}}, \parallel f^- \parallel_{L^{\infty}}\}$, and $\bar{L}$ the maximum value of $L$ over the support of $f^+$ or $f^-$.
\begin{align}\label{eq:3.63}
|j|&=|\frac{\pi}{t^2}\int_{-\infty}^{\infty}\int_{0}^{\infty}w(f^++f^-)dLdw|\nonumber\\
&\leq\frac{4\pi\bar{L}F}{t^2}\bar{w}^2\nonumber\\
&\leq4\pi\bar{L}F C_{1}^{2} t^{-4+2\epsilon},
\end{align}
\begin{align}\label{eq:3.62}
|p|&\leq\frac{\pi}{t^2}\int_{-\infty}^{\infty}\int_{0}^{\infty}\frac{w^2}{\sqrt{1+w^2+L/t^2}}(f^++f^-)dLdw+\frac{1}{2}(e^{2\lambda}E^2+ct^{-4})\nonumber\\
&\leq \frac{\pi}{t^2}\int_{-\infty}^{\infty}\int_{0}^{\infty}|w|(f^++f^-)dLdw+\frac{1}{2}(e^{2\lambda}E^2+ct^{-4})\nonumber\\
&\leq(4\pi\bar{L}F C_{1}^{2}+\frac{1}{2}C_{1}^{2}+\frac{1}{2}c) t^{-4+2\epsilon},
\end{align}

\begin{align}\label{eq:3.64}
|b|&=|\frac{\pi}{t^2}\int_{-\infty}^{\infty}\int_{0}^{\infty}\frac{w}{\sqrt{1+w^2+L/t^2}}(f^+-f^-)dLdw|\nonumber\\
&\leq4\pi\bar{L}F C_{1}^{2} t^{-4+2\epsilon}.
\end{align}
An estimate for $e^{2\mu}$ is also required.
\begin{align*}
e^{2\mu}&=\Lambda^{-1}t^{-2}(\Lambda t^{2}e^{2\mu})\\
&\leq \Lambda^{-1}t^{-2}[(\Lambda t^{2}e^{2\mu}-3-3k e^{2\mu})+3+3k e^{2\mu}]\\
&\leq \Lambda^{-1}t^{-2}[C_1 t^{-3+\epsilon}+3+3k e^{2\mu}].
\end{align*}
If $k\leq0$ the last term in the latter inequality can be discarded. If $k>0$ then we need $3\Lambda^{-1}t^{-2}<1$ i.e. $\Lambda t^2>3$. Assume for the moment that $k\leq0$. Then
\begin{align}\label{eq:3.65}
e^{2\mu}&\leq\Lambda^{-1}t^{-2}[C_1 t^{-3+\epsilon}+3]\nonumber\\
&\leq3\Lambda^{-1}t^{-2}+C_1\Lambda^{-1}t^{-5+\epsilon}.
\end{align}
It follows that
\begin{align*}
|4\pi te^\mu j|&\leq 4\pi t[3\Lambda^{-1}t^{-2}+C_1\Lambda^{-1}t^{-5+\epsilon}]^{1/2}\times4\pi \bar{L}FC_{1}^{2}t^{-4+2\epsilon}\\&\leq16\pi^2\bar{L}FC_{1}^{2}t^{-3+2\epsilon}[\sqrt{3\Lambda^{-1}}t^{-1}+C_{1}^{1/2}\Lambda^{-1/2}t^{-5/2+\epsilon/2}],
\end{align*}
keeping the worst powers and using (\ref{eq:3.53}) implies
\begin{align}\label{eq:3.68}
|e^{-\lambda}\mu'|\leq 16\pi^2\bar{L}FC_{1}^{2}(\sqrt{3\Lambda^{-1}}+C_{1}^{1/2}\Lambda^{-1/2})t^{-4+2\epsilon}=:C_2t^{-4+2\epsilon}.
\end{align}
Note that here there is no dependence on the initial data except for the conserved quantities $\bar{L}$ and $F$. Moreover $C_2=O(C_{1}^{2})$.

From (\ref{eq:3.64})-(\ref{eq:3.65}) we have
\begin{align*}
|\partial_t(t^2e^\lambda E)|&=|t^2e^\mu b|\\
&\leq [3\Lambda^{-1}t^{-2}+C_1\Lambda^{-1}t^{-5+\epsilon}]^{1/2}\times4\pi \bar{L}FC_{1}^{2}t^{-2+2\epsilon},
\end{align*}
keeping the worst powers gives
\begin{align*}
|\partial_t(t^2e^\lambda E)|\leq (\sqrt{\frac{3}{\Lambda}}+C_{1}^ {1/2}\Lambda^{-1/2})\times4\pi \bar{L}FC_{1}^{2}t^{-3+2\epsilon},
\end{align*}
and integrating this in time yields
\begin{align}\label{eq:3.69}
(e^\lambda E)(t)&\leq t_{0}^{2}e^{\lambda(t_0)}|E(t_0)|t^{-2}+\frac{1}{2-2\epsilon}(\sqrt{\frac{3}{\Lambda}}+C_{1}^{1/2}\Lambda^{-1/2})\times4\pi \bar{L}FC_{1}^{2}t^{-4+2\epsilon}\nonumber\\
&\leq\left[t_{0}^{2}e^{\lambda(t_0)}|E(t_0)|+\frac{2\pi}{1-\epsilon}(\sqrt{\frac{3}{\Lambda}}+C_{1}^{1/2}\Lambda^{-1/2}) \bar{L}FC_{1}^{2}\right]t^{-2}=:C_3 t^{-2}.
\end{align}
The constant $C_3$ in the last inequality depends in a transparent way on the initial data.
We have
\begin{align*}
8\pi t^2 p\leq 4\pi(8\pi\bar{L}F C_{1}^{2}+C_{1}^{2}+c) t^{-2+2\epsilon},
\end{align*}
so that using (\ref{eq:3.58}) and integration gives
\begin{align*}
|-\frac{1}{3}t e^{-2\mu}(\Lambda t^2e^{2\mu}-3-3k e^{2\mu})|\leq\frac{1}{3}t_0 e^{-2\mu(t_0)}&|\Lambda t_{0}^{2}e^{2\mu(t_0)}-3-3k e^{2\mu(t_0)}|\\&+|\frac{4\pi}{-1+2\epsilon} (8\pi\bar{L}F C_{1}^{2}+C_{1}^{2}+c) t^{-1+2\epsilon}|.
\end{align*}
At this point the assumption $\epsilon<1/2$ is needed. Then
\begin{equation*}
|-\frac{1}{3}t e^{-2\mu}(\Lambda t^2e^{2\mu}-3-3k e^{2\mu})|\leq\frac{1}{3}t_0 |\Lambda t_{0}^{2}-3e^{-2\mu(t_0)}-3k|+\frac{4\pi}{1-2\epsilon} (8\pi\bar{L}F C_{1}^{2}+C_{1}^{2}+c).
\end{equation*}
Using (\ref{eq:3.65}) and keeping the worst powers, it follows that
\begin{align}\label{eq:3.70}
|\Lambda& t^2e^{2\mu}-3-3k e^{2\mu}|\nonumber\\&\leq[3\Lambda^{-1}+C_1 \Lambda^{-1}]\left[t_0 |\Lambda t_{0}^{2}-3e^{-2\mu(t_0)}-3k|+\frac{12\pi}{1-2\epsilon} (8\pi\bar{L}F C_{1}^{2}+C_{1}^{2}+c)\right]t^{-3}\nonumber\\&=:C_4t^{-3}.
\end{align}
Now let us examine the evolution of $w$.\\Using (\ref{eq:3.65}), (\ref{eq:3.68}), (\ref{eq:3.69}), the bootstrap assumption, the fact that $C_1<1$ and then keeping the worst powers gives
\begin{align*}
|-t e^{\mu}(e^{-\lambda}\mu')&\sqrt{1+w^2+L/t^2}\pm te^{\mu}(e^{\lambda}E)|\\
&\leq \left[3t_{0}^{2}e^{\lambda(t_0)}|E(t_0)|+18\pi\bar{L}F\Lambda^{-1/2}C_{1}^{2}(\frac{1}{1-\epsilon}+8\pi\Lambda^{-1/2}(2+\sqrt{\bar{L}}))\right]t^{-2},
\end{align*}
and using (\ref{eq:3.59}), the bootstrap assumption and integration it follows that
\begin{align}\label{eq:3.71}
t\bar{w}(t)&\leq t_0\bar{w}(t_0)+\int_{t_0}^{t}C_1 s^{-3+\epsilon}s\bar{w}(s)ds\nonumber\\&+\int_{t_0}^{t}\left[3t_{0}^{2}e^{\lambda(t_0)}|E(t_0)|+18\pi\bar{L}F\Lambda^{-1/2}C_{1}^{2}(\frac{1}{1-\epsilon}+8\pi\Lambda^{-1/2}(2+\sqrt{\bar{L}}))\right]s^{-2}ds\nonumber\\&\leq t_0\bar{w}(t_0)+6t_{0}^{2}e^{\lambda(t_0)}|E(t_0)|+36\pi\bar{L}F\Lambda^{-1/2}C_{1}^{2}\left(\frac{1}{1-\epsilon}+8\pi\Lambda^{-1/2}(2+\sqrt{\bar{L}})\right)\nonumber\\&+\frac{C_{1}^{2}}{1-\epsilon}=:C_5.
\end{align}
An estimate on the matter quantity $\rho$ is needed. Using (\ref{eq:3.71}) gives 
\begin{align}\label{eq:3.60}
\rho_{vl}&:=\frac{\pi}{t^2}\int_{-\infty}^{\infty}\int_{0}^{\infty}\sqrt{1+w^2+L/t^2}(f^++f^-)dLdw \nonumber\\
&\leq\frac{\pi}{t^2}\int_{-\bar{w}}^{\bar{w}}\int_{0}^{\bar{L}}\sqrt{1+\bar{w}^2+L/t^2}2FdLdw \nonumber\\
&\leq \frac{4\pi\bar{L}F}{t^2}\bar{w}\sqrt{1+\bar{w}^2+L/t^2}\nonumber\\
&\leq\frac{4\pi\bar{L}F}{t^2}C_5 t^{-1}\sqrt{1+C_{5}^{2} t^{-2}+\bar{L}t^{-2}}\nonumber\\
&\leq4\pi\bar{L}F C_5 t^{-3}(1+C_5t^{-1}+\sqrt{\bar{L}}t^{-1}),
\end{align}
\begin{align}\label{eq:3.61}
\rho&:=\rho_{vl}+\frac{1}{2}(e^\lambda E)^2+\frac{1}{2}ct^{-4}\nonumber\\
&\leq4\pi\bar{L}F C_5 t^{-3}(1+C_5t^{-1}+\sqrt{\bar{L}}t^{-1})+\frac{1}{2}C_{1}^{2}t^{-4+2\epsilon}+\frac{1}{2}ct^{-4}.
\end{align}
In the case of plane symmetry $k=0$ equations (\ref{eq:3.65}), (\ref{eq:3.70}) and (\ref{eq:3.61}) imply that
\begin{align}\label{eq:3.66}
|\frac{1}{2}&(\Lambda e^{2\mu}t^2-3-3k e^{2\mu})+k e^{2\mu}+4\pi t^2 e^{2\mu}\rho|\nonumber\\
&\leq \frac{1}{2}C_4t^{-3}+4\pi t^2[3\Lambda^{-1}t^{-2}+C_1\Lambda^{-1}t^{-5+\epsilon}]\times\nonumber\\
&\left[4\pi\bar{L}F C_5 t^{-3}(1+C_5t^{-1}+\sqrt{\bar{L}}t^{-1})+\frac{1}{2}C_{1}^{2}t^{-4+2\epsilon}+\frac{1}{2}ct^{-4}\right]\nonumber\\
&\leq \left[\frac{1}{2}C_4+16\pi^2 \bar{L}F\Lambda^{-1}C_5(3+C_1)(1+C_5+\sqrt{\bar{L}})+\frac{1}{2}C_{1}^{2}+\frac{1}{2}c\right]t^{-2}.
\end{align}
Whereas in the case of hyperbolic symmetry $k=-1$, $\Lambda^{-1}(3+C_1)t^{-2}$ appears as an adding term in the right hand side of equation (\ref{eq:3.66}), i.e.
\begin{align}\label{eq:3.68}
|\frac{1}{2}&(\Lambda e^{2\mu}t^2-3-3k e^{2\mu})+k e^{2\mu}+4\pi t^2 e^{2\mu}\rho|\nonumber\\
&\leq \left[\frac{1}{2}C_4+16\pi^2 \bar{L}F\Lambda^{-1}C_5(3+C_1)(1+C_5+\sqrt{\bar{L}})+\frac{1}{2}C_{1}^{2}+\frac{1}{2}c+\Lambda^{-1}(3+C_1)\right]t^{-2}.
\end{align}

For the latter inequalities we only kept the worst powers. We have
\begin{equation}\label{eq:3.67}
|t\dot\lambda-1|\leq C_6t^{-2},
\end{equation}
where $C_6$ is the constant written out in (\ref{eq:3.66}) for the case $k=0$, and in (\ref{eq:3.68}) for the case $k=-1$.

The constants $C_2$-$C_6$ appearing in equations (\ref{eq:3.71})-(\ref{eq:3.67}) are all less than or equal to $C\times(g(\delta)+C_{1}^{2})$, with $C$ a positive constant and $g(\delta)$ a positive function of $\delta$ tending to $0$ as $\delta$ tends to $0$. Therefore it is always possible to choose $C_1$ and $\delta$ small enough in such a way that $CC_1\leq1/2$ and $Cg(\delta)\leq C_1/2$, and so the constants $C_2$-$C_6$ are all less than $C_1$. This closes the bootstrap argument as it implies that (\ref{boot1})-(\ref{boot2}) hold  on an interval $[t_0,t_1)$, with $t_1>t_*$. This contradicts the maximality of the interval $[t_0,t_*)$. Therefore $t_*=\infty$. To complete the proof of theorem \ref{decay} it remains to show that the spacetime is complete. In fact recall that as in \cite{tchapnda1} the relation between coordinate time $t$ and proper time $\tau$ along the trajectory is given by
\[
\frac{d\tau}{dt}=\frac{e^\mu}{\sqrt{m^2+w^2+L/t^2}}.
\]
The decay estimate on $e^\lambda E$ and equations (\ref{eq:2.8}) and (\ref{eq:1.11}) can be used to obtain the inequality
\begin{equation*}
e^{2\mu}\geq \frac{t}{C+(C-k)t+\frac{\Lambda}{3}t^3}, \ {\rm for} \ k\leq0.
\end{equation*}
It follows that
\[
e^\mu\geq Ct^{-1}, \ t\geq t_0.
\]
Thus
\[
\frac{d\tau}{dt}\geq \frac{Ct^{-1}}{\sqrt{m^2+C+L}},
\]
and so $\tau$ goes to infinity as does $t$. Theorem \ref{decay} is then proved.
\vskip 10pt\noindent \textbf{Acknowledgement} : The author thanks A.D. Rendall for fruitful suggestions, and the anonymous referee for constructive criticisms. 

\author{Sophonie Blaise Tchapnda\\
Max Planck Institute for Gravitational Physics\\
Albert Einstein Institute\\
Am M\"uhlenberg 1, D-14476 Golm, Germany\\
\texttt{email: tchapnda@aei.mpg.de}\\
On leave from: {\it Department of Mathematics, Faculty of Science\\
University of Yaounde I, PO Box 812, Yaounde, Cameroon\\
\texttt{email: tchapnda@uycdc.uninet.cm}}}
\end{document}